\documentclass[reprint,prl,aps, floatfix]{revtex4-1}
\usepackage{graphicx}
\usepackage{xcolor}
\usepackage{dcolumn}
\usepackage{bm}
\usepackage{amssymb}
\usepackage{braket}
\usepackage{color,amsmath}
\usepackage{comment}
\usepackage{gensymb}
\usepackage{soul,xcolor} 
\setstcolor{red} 
\usepackage{epstopdf}
\usepackage{hhline}
\usepackage{tabularx}
\usepackage{xspace}
\usepackage{layouts}
\usepackage{lineno} 

\newcolumntype{C}[1]{>{\centering\arraybackslash}m{#1}}
\newcolumntype{N}{@{}m{0pt}@{}}

\definecolor{cadmiumgreen}{rgb}{0.0, 0.42, 0.24}

\newcommand{\moire}{moir\'e\xspace}

\begin{document}

\title{Topological flat bands in a family of multilayer graphene moir\'e lattices}

\author{Dacen Waters$^{1,2*}$}
\author{Ruiheng Su$^{3,4*}$}
\author{Ellis Thompson$^{1*}$}
\author{Anna Okounkova$^{1}$}
\author{Esmeralda Arreguin-Martinez$^{5}$}
\author{Minhao He$^{1, 6}$}
\author{Katherine Hinds$^{5}$}
\author{Kenji Watanabe$^{7}$} 
\author{Takashi Taniguchi$^{8}$} 
\author{Xiaodong Xu$^{1,5}$}
\author{Ya-Hui Zhang$^{9}$}
\author{Joshua Folk$^{3,4\dagger}$}
\author{Matthew Yankowitz$^{1,5\dagger}$}

\affiliation{$^{1}$Department of Physics, University of Washington, Seattle, Washington, 98195, USA}
\affiliation{$^{2}$ Intelligence Community Postdoctoral Research Fellowship Program, University of
Washington, Seattle, Washington, 98195, USA}
\affiliation{$^{3}$ Stewart Blusson Quantum Matter Institute, University of British Columbia, Vancouver, British Columbia, V6T 1Z1, Canada}
\affiliation{$^{4}$ Department of Physics and
Astronomy, University of British Columbia, Vancouver, British Columbia, V6T 1Z1, Canada}
\affiliation{$^{5}$Department of Materials Science and Engineering, University of Washington, Seattle, Washington, 98195, USA}
\affiliation{$^{6}$Department of Physics, Princeton University, Princeton, New Jersey, 08544, USA}
\affiliation{$^{7}$Research Center for Functional Materials,
National Institute for Materials Science, 1-1 Namiki, Tsukuba 305-0044, Japan}
\affiliation{$^{8}$International Center for Materials Nanoarchitectonics,
National Institute for Materials Science, 1-1 Namiki, Tsukuba 305-0044, Japan}
\affiliation{$^{9}$Department of Physics and Astronomy, Johns Hopkins University, Baltimore, Maryland, 21205, USA}
\affiliation{$^{*}$These authors contributed equally to this work.}
\affiliation{$^{\dagger}$ jfolk@physics.ubc.ca (J.F.); myank@uw.edu (M.Y.)}

\maketitle

\textbf{Moir\'e materials host a wealth of intertwined correlated and topological states of matter, all arising from flat electronic bands with nontrivial quantum geometry \cite{Balents2020, Andrei2020}. A prominent example is the family of alternating-twist magic-angle graphene stacks, which exhibit symmetry-broken states at rational fillings of the moir\'e band and superconductivity close to half filling \cite{Park2021, Hao2021, Park2022, Burg2022, Zhang2022}. Here, we introduce a second family of twisted graphene multilayers made up of twisted sheets of $M$- and $N$-layer Bernal-stacked graphene flakes. Calculations indicate that applying an electric displacement field isolates a flat and topological moir\'e conduction band that is primarily localized to a single graphene sheet below the moir\'e interface. Phenomenologically, the result is a striking similarity in the hierarchies of symmetry-broken phases across this family of twisted graphene multilayers. Our results show that this family of structures offers promising new opportunities for the discovery of exotic new correlated and topological phenomena, enabled by using the layer number to fine tune the flat moir\'e band and its screening environment.}

Twisting two monolayer graphene sheets by an angle of $\theta\approx1.1^{\circ}$ creates magic-angle twisted bilayer graphene (MATBG), in which several new phases of matter have been realized~\cite{Balents2020, Andrei2020, Bistritzer2011, SuarezMorell, Cao2018a, Cao2018b, Lu2019, Yankowitz2019}. A much broader range of novel physics can be unveiled in closely related structures having three or more sheets of graphene. Early experiments investigated the strongly correlated and topological physics arising in twisted monolayer-bilayer and double-bilayer graphene, noting intriguing similarities between the two systems that were, nevertheless, qualitatively distinct from MATBG~\cite{Chen2021,Polshyn2020,Shi2021,He2021tmbg,Burg2019,Shen2020,Liu2020,Cao2020,He2021tdbg,Kuiri2022, Liu2022, Liu2023}. Whereas the investigation of MATBG has expanded to alternating-twist structures up to five layers~\cite{Park2021,Hao2021,Park2022,Burg2022,Zhang2022}, the study of structures that include Bernal-stacked components has until now been limited to just those two, despite many others in the twisted $M+N$ family carrying predictions of closely related flat bands (t$M+N$, where $M$ and $N$ are positive integers representing the number of Bernal-stacked graphene layers twisted atop one another)~\cite{Goodwin2021}.

From a symmetry perspective, t$M+N$ structures differ fundamentally from the family of alternating-twist magic-angle graphene stacks by the breaking of $C_{2z}$ symmetry (in-plane rotation by 180$^{\circ}$). This allows a gap to be opened between the lowest valence and conduction bands at charge neutrality by an electric displacement field, $D$. The collective Berry curvature of many graphene layers contributes to a non-zero valley Chern number of the moir\'e bands, yielding topological electronic states when interactions generate a spontaneous valley polarization. Experimentally, an intriguing result is that the correlated electronic states appearing in t$1+2$ are rich with emergent topology~\cite{Chen2021,Polshyn2020,Shi2021,He2021tmbg}, and are furthermore remarkably similar to many of the states seen in t$2+2$~\cite{Shen2020,Liu2020,Cao2020,He2021tdbg,Kuiri2022, Liu2022, Liu2023}. The intertwined correlated and topological states are most similar when the direction of $D$ is oriented from the monolayer to the bilayer of t$1+2$, which localizes the conduction band mostly to the bilayer side.

\begin{figure*}[t]
\includegraphics[width=\textwidth]{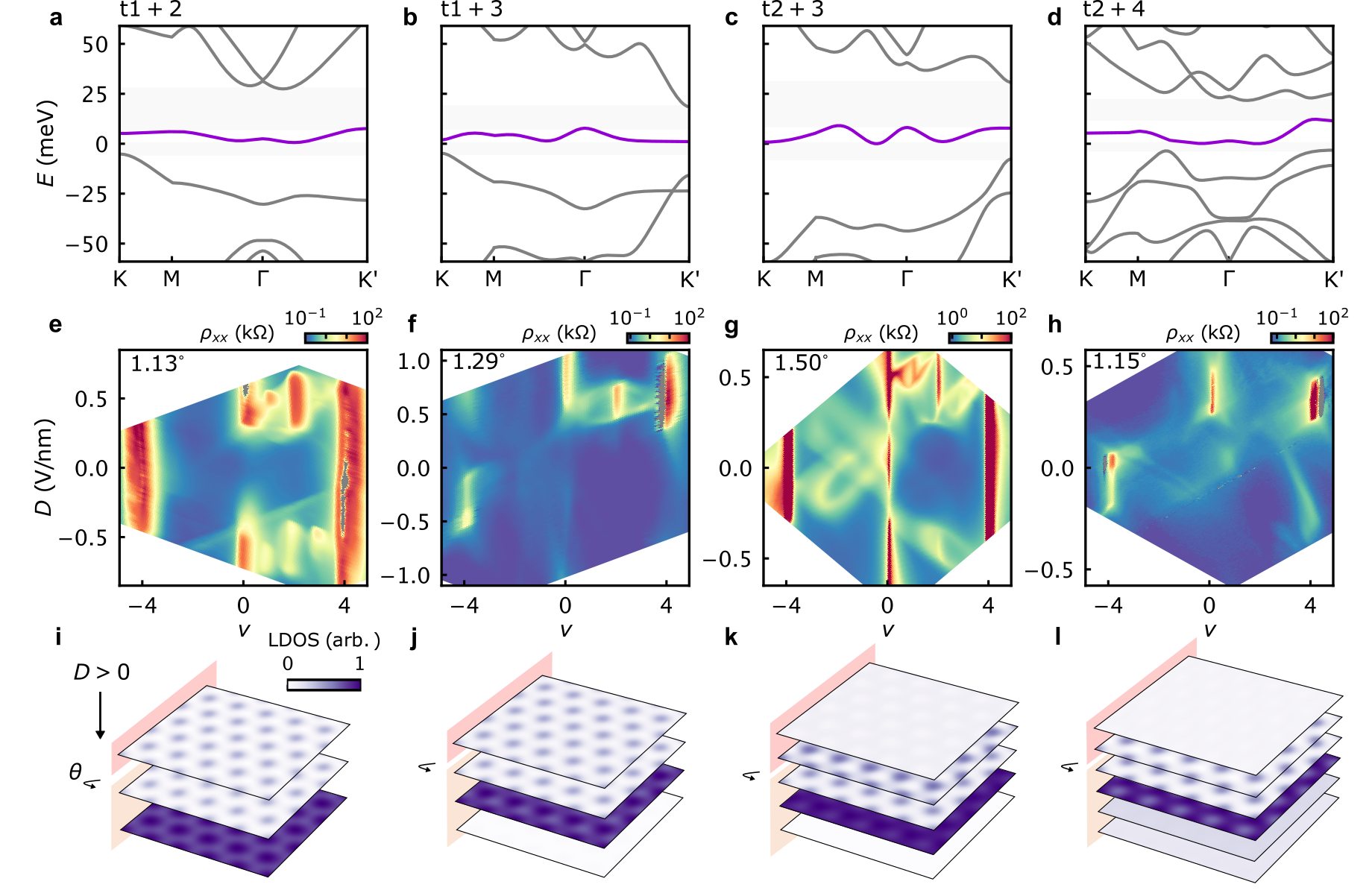} 
\caption{\textbf{Non-interacting features of t$\bm{M+N}$ graphene.}
\textbf{a}, Band structure calculations for t$1+2$ at its optimal twist angle ($\theta=1.13^{\circ}$) and interlayer potential ($\delta=50 \rm{meV}$); defined as when the \moire conduction band (purple curve) has the narrowest dispersion while being simultaneously gapped (grey shaded regions) from both neighboring bands. Energy is measured with respect to the bottom of the \moire conduction band. 
\textbf{b-d}, Similar calculations for (\textbf{b}) t$1+3$ ($\theta = 1.3^{\circ}, \delta = 50 \rm{meV}$); (\textbf{c}) t$2+3$ ($\theta=1.45^{\circ}, \delta = 90 \rm{meV}$); and (\textbf{d}) t$2+4$ ($\theta=1.15^{\circ}, \delta = 50 \rm{meV}$).
\textbf{e}, Resistivity at zero magnetic field as a function of $\nu$ and $D$ in a t$1+2$ device at $\theta=1.13^\circ$. Gray color denotes experimental artifacts where negative resistance is observed, attributed to the effects of highly resistive states. 
\textbf{f-h}, Similar measurements for (\textbf{f}) t$1+3$ with $\theta=1.29^{\circ}$; (\textbf{g}) t$2+3$ with $\theta=1.50^{\circ}$; and (\textbf{h}) t$2+4$ with $\theta=1.15^{\circ}$ . Positive $D$ is defined as pointing from the thin component to the thick component for all systems. Measurements taken at $T=1.7\ \rm{K}$, except for \textbf{e}, where $T=0.3\ \rm{K}$.
\textbf{i-l}, Layer-resolved LDOS calculated for the \moire conduction band, corresponding to the layer combinations and parameters in panels \textbf{(a-d)}. Red and orange shadings delineate the $M$ and $N$ layers above and below the twisted interface. The interlayer potential used in the calculations corresponds to positive $D$ in the experiment. 
}
\label{fig:1}
\end{figure*}

In this work, we extend the study of t$M+N$ graphene structures to include configurations as thick as six total layers, focusing on t$1+3$, t$2+3$, t$1+4$, and t$2+4$. Despite the inclusion of these additional layers of graphene in the moir\'e structure, we find striking commonalities in both the non-interacting and correlated physics across this entire family. Continuum model calculations indicate that these unexpected similarities likely arise because low-energy states in the conduction band are localized to just three graphene sheets, irrespective of the total number of layers in the structure: the two layers at the twisted interface and one more immediately adjacent. Ultimately, the result is that adding graphene sheets above and below does not significantly affect the band structure, simply protecting the trio of active layers and strengthening the roles of topology and correlations in shaping the electronic system.

\begin{figure*}[t]
\includegraphics[width=\textwidth]{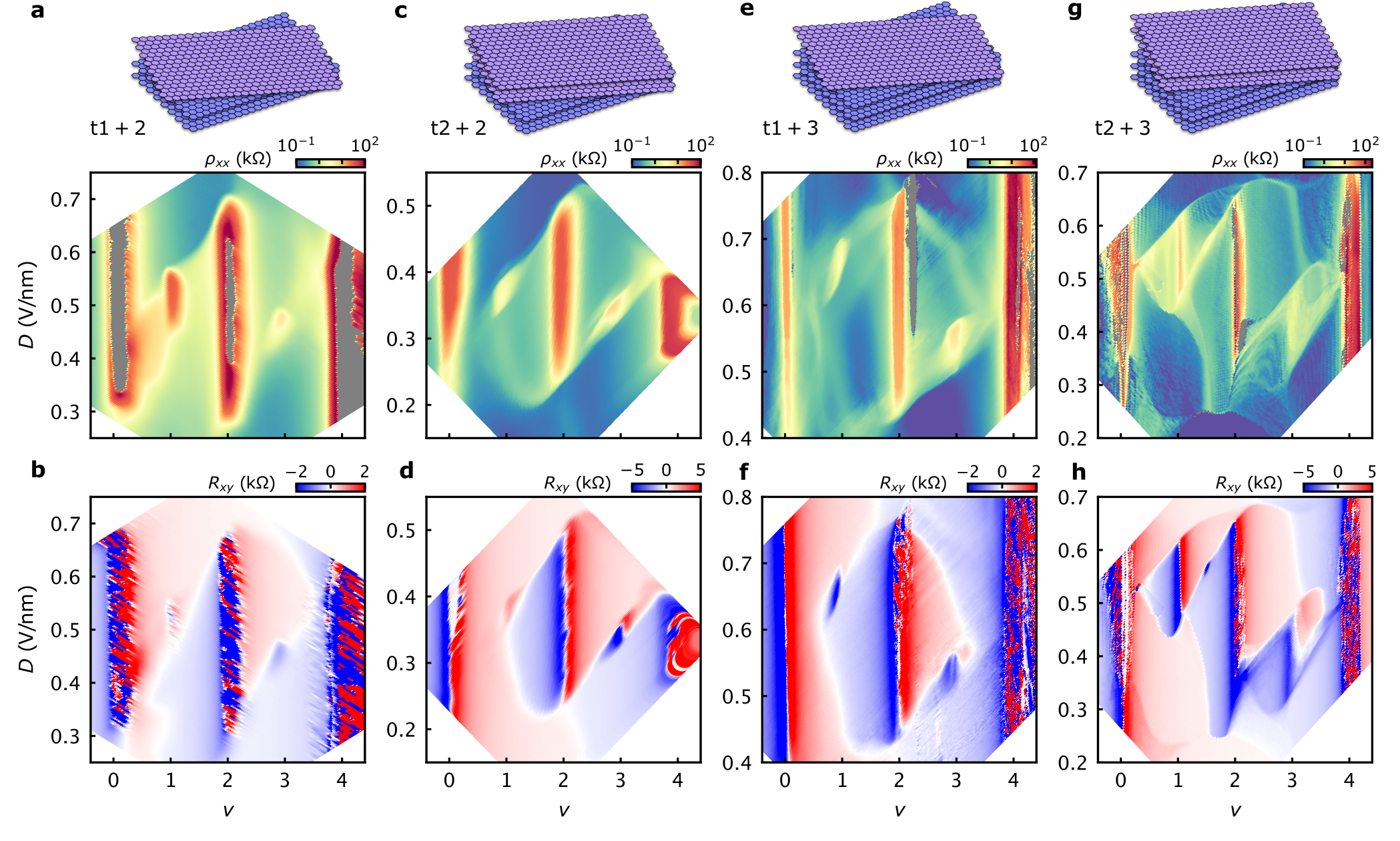} 
\caption{\textbf{Symmetry-broken states in various t$\bm{M+N}$ systems.}
\textbf{a}, Map of the longitudinal resistivity around the correlated states in the t$1+2$ device at $\theta=1.13^{\circ}$. The map is symmetrized at $B=\pm0.5$~T.
\textbf{b}, Similar map of the Hall resistance, anti-symmetrized at $B=\pm0.5$~T.
\textbf{c,e,g}, Analogous $\rho_{xx}$ maps for the \textbf{(c)} t$2+2$ device with $\theta=1.30^\circ$; \textbf{(e)} t$1+3$ device with $\theta=1.29^\circ$; and t$2+3$ device with $\theta=1.50^\circ$.
\textbf{d},\textbf{f},\textbf{h}, Analogous $R_{xy}$ maps for the same set of devices.
Schematics in the top panels indicate the layer combination for each measurement. The measurements are performed at nominal sample temperatures of \textbf{(a-b)} $T=300$~mK, \textbf{(c-f)} $T=20$~mK, \textbf{(g-h)} $T=100$~mK.
}
\label{fig:2}
\end{figure*}

\medskip\noindent\textbf{Isolated moir\'e-localized flat bands}

We first compare the band structures of various representative t$M+N$ structures (t$1+2$, t$1+3$, t$2+3$, t$2+4$, Figs.~\ref{fig:1}a-d) as predicted by the Bistritzer-MacDonald continuum model~\cite{Bistritzer2011} (see Methods). Upon incorporating an interlayer potential arising from an external $D$ pointing from the thinner to the thicker graphene constituent, the moir\'e conduction band in each of these structures (purple in Figs.~\ref{fig:1}a-d) becomes relatively flat and isolated by gaps to both the moir\'e valence band and higher moir\'e conduction band. Other gaps between neighboring bands can also open, with the details depending sensitively on the values of both $\theta$ and $D$ for each particular t$M+N$ construction.

Band structure predictions for each of these structures are corroborated by maps of the longitudinal resistivity, $\rho_{xx}$. The data are collected as a function of top- and back-gate voltages, shown in Figs.~\ref{fig:1}e-h after converting the gate voltages to moir\'e band filling, $\nu$, and displacement field, $D$ (see Methods). In qualitative agreement with the band structure calculations, all exhibit insulating states at the charge neutrality point ($\nu=0$) and at full-filling of the lowest moir\'e valence and conduction bands ($\nu=\pm4$) over certain ranges of $D$, marked by large values of $\rho_{xx}$ that exceed $h/e^2$ and diverge as the temperature is lowered ($h$ is Planck's constant and $e$ is the charge of the electron). 

It is not immediately obvious that isolated flat bands would form in many of these t$M+N$ structures. Bernal graphene films with three or more layers feature multiple low-energy bands~\cite{Partoens2006, Waters2023}, all of which must hybridize with the bands from the other twisted constituent to yield an isolated moir\'e band. To help explain how different t$M+N$ constructions form similar moir\'e flat bands, Figs.~\ref{fig:1}i-l show calculations of the layer-resolved local density of states (LDOS) at full-filling of the lowest moir\'e conduction band (i.e., integrated over the purple bands in Figs.~\ref{fig:1}a-d).

\begin{figure*}[t]
\includegraphics[width=\textwidth]{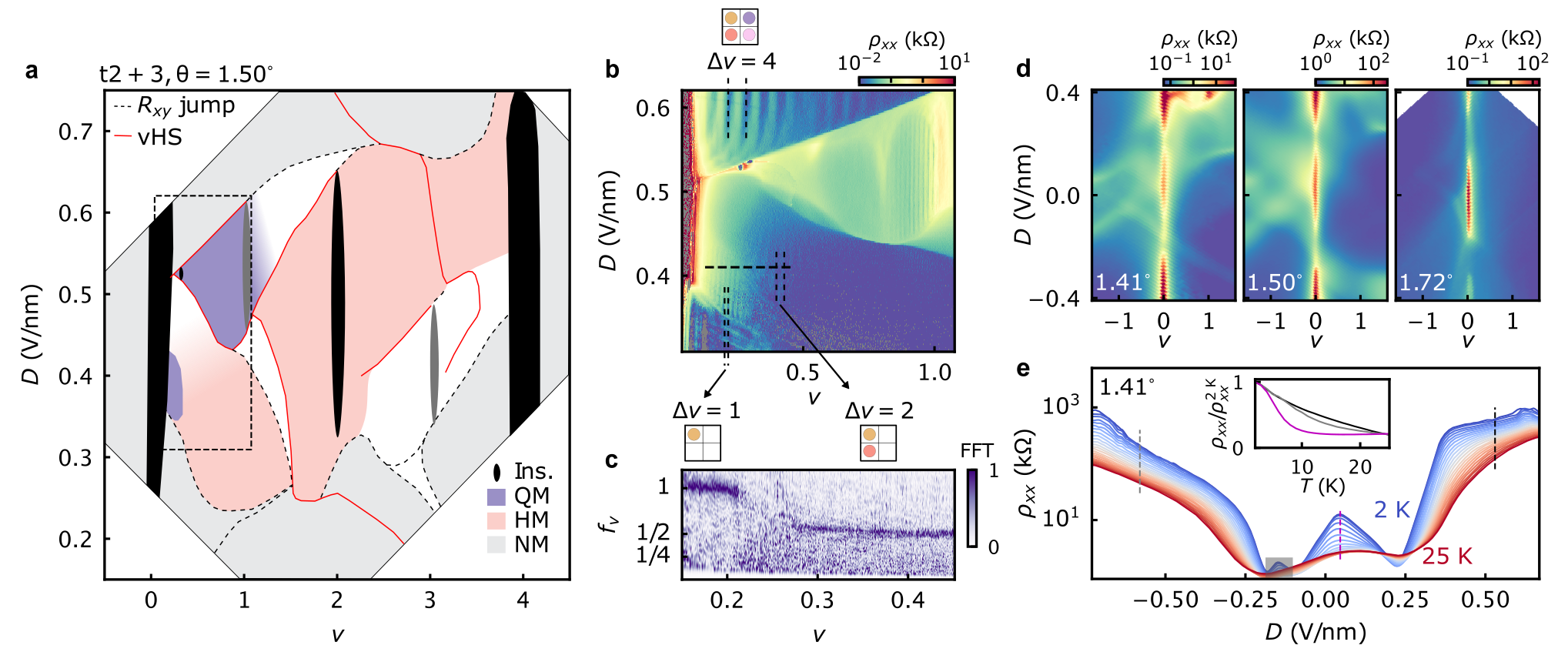} 
\caption{\textbf{Correlated states in t$\mathbf{2+3}$ graphene.}
\textbf{a}, Summary phase diagram determined from magnetotransport measurements in the $\theta=1.50^\circ$ device (see Methods for the determination of different features). Black vertical features indicate insulating states (Ins). The $\nu=1$ and $3$ states are in shaded dark grey to indicate that it is a weakly-developed resistive state corresponding to an incipient insulator. Shaded regions indicate where the metallic states are either normal-metals (NM), half-metals (HM), or quarter-metals (QM). White regions within the bounding box indicate situations in which the degeneracy cannot be uniquely determined. Solid red lines denote likely van Hove singularities (vHS) with $R_{xy}$=0. Dashed black lines denote abrupt jumps in $R_{xy}$. 
\textbf{b}, Map of $\rho_{xx}$ acquired at $T=20$~mK and symmetrized at $B\pm0.9$~T, corresponding to the black dashed box in \textbf{(a)}. Spacing of the quantum oscillations ($\Delta \nu$) indicates the degeneracy of the metallic phases. Corresponding schematics indicate the degeneracy for representative regions of normal-metal ($\Delta \nu=4$, top), half-metal ($\Delta \nu=2$, bottom right), and quarter-metal ($\Delta \nu=1$, bottom left) phases. 
\textbf{c}, Fourier transform analysis (FFT) of the SdH measurements taken along the dashed black line at $D=0.41$~V/nm in \textbf{(b)}. The FFT amplitude is normalized to the maximum value of the measurement, and the frequency of oscillations is normalized to the density (see Methods for details).
\textbf{d}, Measurements of $\rho_{xx}$ for three t$2+3$ devices with different $\theta$ acquired around $\nu=0$. All measurements are performed at $T=2$~K. 
\textbf{e}, Temperature dependence of $\rho_{xx}$ as a function of $D$, acquired in the $\theta=1.41^\circ$ device at $\nu=0$. Line cuts at select values of $D$ are shown for each state in the inset, normalized by their respective values at $T=2$~K. The gray shaded region indicates a region of the data influenced by artifacts from the electrical contacts. 
}
\label{fig:3}
\end{figure*}

Considering first the t$1+2$ structure (Fig.~\ref{fig:1}i), the moiré potential localizes the LDOS on a triangular lattice of ABB-stacked sites, with most weight appearing on the graphene sheet one below the moir\'e interface. The layer-resolved LDOS configuration of t$1+2$ recurs in the thicker t$M+N$ structures, with the additional layers of graphene away from the twisted interface carrying a comparatively small density of states. In this sense, the t$1+2$ structure can be considered as the basic building block of all of the thicker t$M+N$ constructions. Given that their low-energy bands are all similarly localized nearby the moir\'e interface, it is natural to expect that the physics of all of these t$M+N$ structures may exhibit common features. 

\medskip\noindent\textbf{Common features of the correlated phases}

The transport measurements in Figs.~\ref{fig:1}e-h exhibit insulating states at certain integer values of $\nu$ beyond those predicted by the single-particle band structure. As previously observed in a variety of other moir\'e systems, these correlated insulators arise as a consequence of spontaneous symmetry breaking within the moir\'e flat bands due to Coulomb interactions~\cite{Balents2020, Andrei2020, Bistritzer2011, SuarezMorell, Cao2018a, Cao2018b, Lu2019, Yankowitz2019, Chen2021,Polshyn2020,Shi2021,He2021tdbg,He2021tmbg,Liu2020,Cao2020,Burg2019,Shen2020,Kuiri2022, Park2021, Hao2021, Park2022, Burg2022, Zhang2022, Liu2022, Liu2023}. We now turn our attention to analyzing the properties of the correlated phases across the family of t$M+N$ structures, comparing to what is known about the symmetry-broken states in t$1+2$ and t$2+2$. 

Figure~\ref{fig:2} shows high-resolution zoom-ins of both longitudinal and Hall resistances, $\rho_{xx}$ and $R_{xy}$, for t$1+2$, t$2+2$, t$1+3$, and t$2+3$, along with simple schematics of the material structure. As was seen for the non-interacting features in Figs.~\ref{fig:1}e-h, the qualitative arrangement of correlated states are similar across all of these t$M+N$ constructions. The most robust insulating state arises at $\nu=2$ in all, spanning the largest range of $D$ and exhibiting the largest value of $\rho_{xx}$ at low temperature. $R_{xy}$ reverses sign across the $\nu=2$ state, consistent with an interaction-induced band gap with hole-like carriers at $\nu\lesssim 2$ and electron-like carriers at $\nu\gtrsim 2$. Previous measurements of the evolution of the $\nu=2$ states in t$1+2$ and t$2+2$ with in-plane magnetic field indicate that they are likely spin polarized~\cite{Liu2020,Chen2021,Shen2020,Cao2020}; analogous measurements in our t$1+3$ and t$2+3$ samples are also consistent with spin-polarized insulators (see Extended Data Fig.~\ref{fig:spin-polarization}). All samples exhibited a region of $\nu$ and $D$ surrounding the insulator at $\nu=2$ characterized by a slight increase in $\rho_{xx}$ and an abrupt sign reversal in $R_{xy}$. These features have previously been explained for t$1+2$ and t$2+2$ as arising due to the formation of a spin-polarized half-metal state with a reduced isospin degeneracy of two.

Resistive states additionally appear at $\nu=1$ and $3$ in each of the $\rho_{xx}$ maps in Fig.~\ref{fig:2}. These feature additional sign reversals (or large enhancements) in $R_{xy}$, indicating the formation of additional symmetry-broken states with no remaining isospin degeneracies. Previous studies of t$1+2$ and t$2+2$ suggest that the ground-state ordering of these odd $\nu$ phases is less consistent, as spin-valley polarized (SVP) states compete closely with intervalley coherent (IVC) states~\cite{Chen2021,Polshyn2020,Shi2021,He2021tmbg,Shen2020,Liu2020,Cao2020,He2021tdbg,Kuiri2022}. These two can be challenging to distinguish in an experiment, as we discuss in more detail in the Methods.

\begin{figure*}[t]
\includegraphics[width=\textwidth]{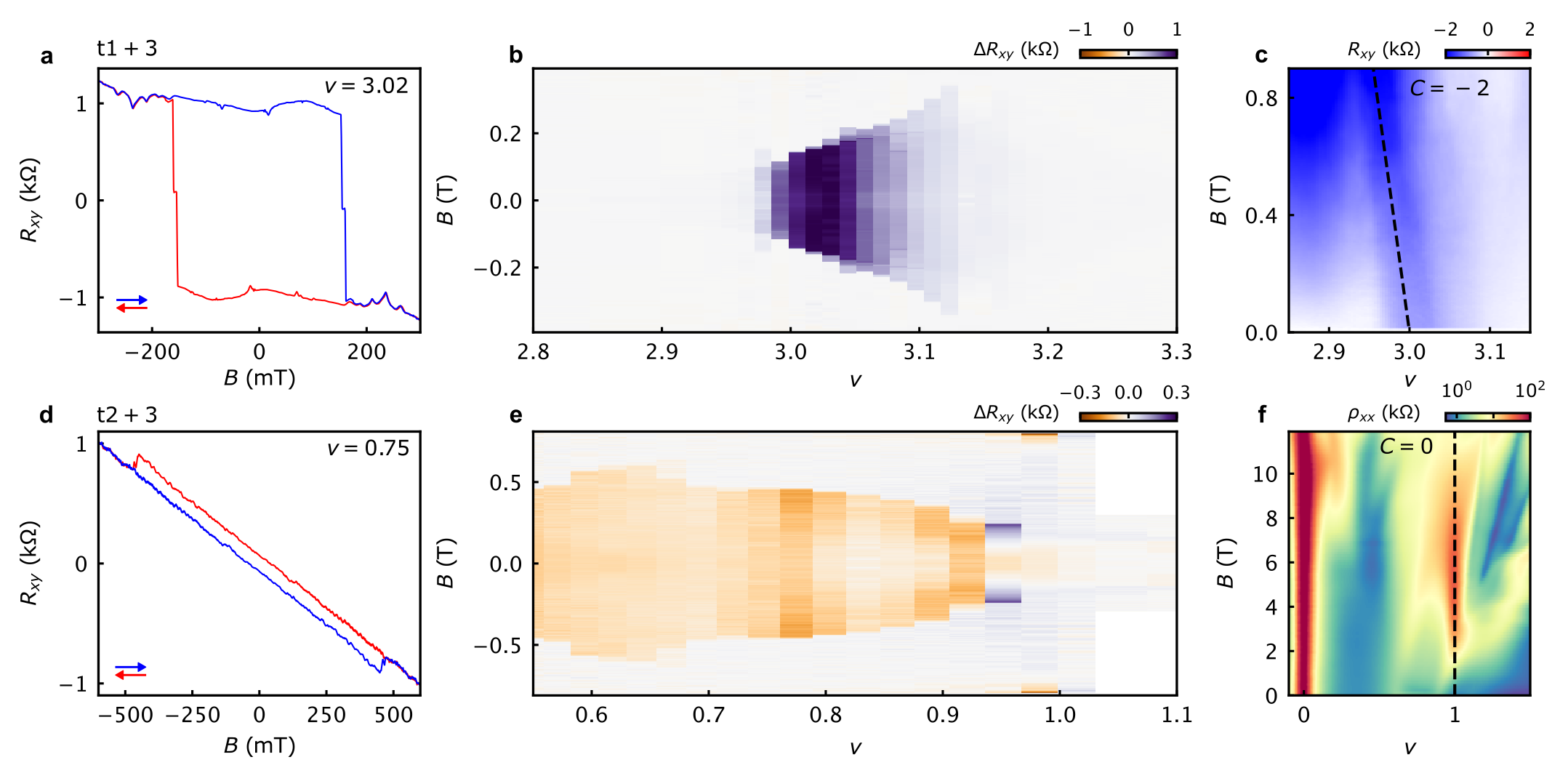} 
\caption{\textbf{Anomalous Hall effects in t$\mathbf{1+3}$ and t$\mathbf{2+3}$ devices.}
\textbf{a}, $R_{xy}$ measurement in the t$1+3$ device ($\theta =1.29^\circ$) acquired as $B$ is swept back and forth at $\nu=3.02$ and $D=0.523$~V/nm.
\textbf{b}, Doping dependence of the AHE effect at the same displacement field characterized by the difference in the forward and backward sweeps, $\Delta R_{xy}=(R_{xy}^{\uparrow}-R_{xy}^{\downarrow})/2$.
\textbf{c}, Landau fan diagram of $R_{xy}$ versus $B$ around $\nu\approx3$. The correlated Chern insulator emerging from $\nu=3$ exhibits a slope consistent with $C=-2$. 
\textbf{d}, $R_{xy}$ measurement in the t$2+3$ device ($\theta =1.50^\circ$) acquired at $\nu=0.75$ and $D=0.532$~V/nm.
\textbf{e}, Doping dependence of the AHE in the same device.
\textbf{f}, Landau fan diagram of $\rho_{xx}$ versus $B$ around $\nu\approx1$. The correlated state at $\nu=1$ projects vertically, consistent with $C=0$. All data acquired at $T=20~\rm{mK}$.
}
\label{fig:4}
\end{figure*}

It is interesting to note that the t$2+4$ device from Figs.~\ref{fig:1}d,h,l, as well as the t$1+4$ and t$2+5$ devices we have studied, did not show correlated insulating states or pronounced regions of enhanced resistivity with an $R_{xy}$ sign reversal at a small magnetic field (Extended Data Fig. \ref{fig:additional_devices} and Supplementary Fig. 2). This may be due to: (i) the smaller band gaps to higher moir\'e bands in these thicker structures, (ii) bands that are more dispersive in reality than those predicted by calculations, (iii) devices made away from the optimal flat-band twist angle for each layer number construction, or some combination of all of these. Indications of a symmetry-broken Fermi surface at $\nu=2$ in t$2+4$ did appear over a small range of magnetic field around $B\approx4$~T (see Extended Data Fig.~\ref{fig:2-4_LF}). However, this correlated phase is quickly suppressed by competing quantum Hall states originating from $\nu=0$. Nevertheless, the new Fermi surface formed at $\nu=2$ in a modest magnetic field indicates that this sample is also close to a strongly correlated regime.

\medskip\noindent\textbf{Abundance of correlated states in t$\mathbf{2+3}$}

Among all the t$M+N$ samples we have studied, t$2+3$ devices had the largest extent of symmetry-broken phases as a function of $\nu$ and $D$ (Fig.~\ref{fig:3}, see also Extended Data Fig.~\ref{fig:2-3_twist_dep}). In addition to exhibiting symmetry-broken correlated insulators for both signs of $D$ in the lowest moir\'e conduction band, the samples showed evidence for symmetry broken phases distinct from the usual states seen in the other t$M+N$ systems (see Fig.~\ref{fig:1}g, Extended Data Fig.~\ref{fig:2-3_high_field_correlations}, and Supplementary Fig. 3). 

Figure~\ref{fig:3}a summarizes our experimental observations for a t$2+3$ device with a 1.50$^{\circ}$ twist angle, incorporating observations from Figs.~\ref{fig:2}g,h,~\ref{fig:3}b,c, and Extended Data Fig.~\ref{fig:2-3_lowfield_QOs}. States labeled in black or dark gray are insulating or highly resistive (the unexpected gapped state at $\nu=0.25$ will be discussed in future work). Solid red curves denote sign changes in $R_{xy}$ that likely correspond to van Hove singularities, whereas dashed black curves denote abrupt jumps in $R_{xy}$ without a sign change, possibly indicating the formation of a new interaction-induced Fermi surface. Shaded regions in Fig.~\ref{fig:3}a correspond to metallic states with different isospin degeneracies. Experimentally, the degeneracy can be identified either from the spacing, $\Delta\nu$, between Shubnikov de Haas (SdH) oscillations in $\nu-D$ maps taken at finite magnetic field (Figs.~\ref{fig:2}g and~\ref{fig:3}b) or from the Fourier transform of SdH oscillations collected by sweeping the magnetic field at fixed gate voltage (Fig.~\ref{fig:3}c and~Extended Data~\ref{fig:2-3_lowfield_QOs}). The degeneracy extracted from $\Delta\nu$ is four outside the correlated region (e.g., $\Delta \nu=4$ in the top of the map shown Fig.~\ref{fig:3}b), consistent with the four-fold spin and valley degeneracy of graphene. The degeneracy is reduced to two inside the region surrounding the correlated insulator at $\nu=2$ (light red in Fig.~\ref{fig:3}a), consistent with our inference of a spin-polarized ground state.

The degeneracy is harder to discern in the regions surrounding $\nu=1$ and $3$ (see Methods and Extended Data Fig.~\ref{fig:2-3-multiband transport}), although there is a region of $\nu$ closely surrounding $\nu=1$, colored in purple, where $\Delta\nu$ is unambiguously one. There is also a small pocket of a symmetry-broken phase over a narrow range of $D\approx0.4$~V/nm between $\nu=0$ and $\nu\approx0.21$. Figure~\ref{fig:3}c shows the normalized Fourier transform of the SdH oscillations measured at a fixed $D$ cutting through this small pocket, as denoted by the horizontal black dashed line in Fig.~\ref{fig:2}b (see Methods for further description of the analysis). We see that all isospin degeneracies are lifted in this pocket, corresponding to a quarter-metal phase for $\nu<0.21$. This is, to our knowledge, the first observation of a quarter-metal state in the t$M+N$ family that is not directly associated with the $\nu=1$ insulator, pointing to the unusually strong interactions in this system. Interestingly, it does not exhibit the anomalous Hall effect (AHE), suggesting that it may either carry a very small Berry curvature or instead be in an IVC ordered state (Extended Data~\ref{fig:2-3_noAHE}).

We further see an unexpected insulating state at $\nu=0$ near $D=0$ across several t$2+3$ devices with different twist angles (Fig.~\ref{fig:3}d and Extended Data Fig.~\ref{fig:2-3_twist_dep}). Our single-particle band structure calculations do not predict such a gap for any reasonable model parameters (see Methods), suggesting that it may arise spontaneously. This hypothesis is supported by measurements of the temperature dependence at $\nu=0$ for the device with $\theta=1.41^\circ$ (Fig.~\ref{fig:3}e). The sample resistance at $|D|>0.3$ steadily increases as the temperature is lowered (grey and black curves in the inset), consistent with a band insulator. Near $D=0$, on the other hand, insulating behavior abruptly onsets below $T\leq 10\ \rm{K}$ (purple curve in the inset). Although not definitive, the abrupt emergence of the gapped state with temperature at $D\approx 0$ suggests that it may arise owing to interactions, similarly to the correlated insulators at charge neutrality in bilayer graphene~\cite{Martin2010,Weitz2010,Mayorov2011,Velasco2012} and rhombohedral few-layer graphene~\cite{Bao2011,Shi2020,Liu2024,Han2024}. Overall, it remains an open question as to why t$2+3$ exhibits the most robust and prevalent symmetry-broken phases over a wide range of $\nu$ and $D$, as our band structure modeling does not predict that the bandwidth should be small compared to other t$M+N$ structures. 
 
\medskip\noindent\textbf{Topological states in t$\bm{M+N}$ graphene}

Finally, we turn our attention to the topological properties of the t$M+N$ moir\'e bands. The AHE has been seen previously in both t$1+2$ and t$2+2$~\cite{Chen2021,Polshyn2020,He2021tmbg,Kuiri2022}. In the former, anomalous Hall resistances close to the quantized values $h/2e^2$ and $h/e^2$ have been observed, consistent with a SVP state formed from bands with a $\theta$-dependent valley Chern number of either $C_v=1$ or $2$~\cite{Polshyn2020,Chen2021}. The AHE is typically not observed in t$2+2$ at $\nu=1$ and $3$, despite strong indications of a valley Chern number of 2, potentially indicating IVC order~\cite{Shen2020,Cao2020,Liu2020,He2021tdbg}. Nevertheless, in select cases it has been observed for $\nu>3$, pointing to the emergence of a SVP state~\cite{Kuiri2022}. 

In our t$1+3$ sample, the correlated state at $\nu=1$ is too weakly developed to determine its isospin ordering (even with an applied magnetic field, as shown in Supplementary Fig. 4). However, the state at $\nu=3$ exhibits a clear AHE characterized by hysteresis in $R_{xy}$ upon sweeping $B$ back and forth, as shown in Fig.~\ref{fig:4}a at optimal doping and displacement field. Figure~\ref{fig:4}b shows the doping dependence of the AHE, characterized by the difference between forward and backward sweeps in a magnetic field, $\Delta R_{xy}=(R_{xy}^{\uparrow}-R_{xy}^{\downarrow})/2$. The AHE state is most pronounced close to $\nu=3$ and is quickly suppressed upon doping. The corresponding Landau fan diagram shows that the gapped state drifts to smaller $\nu$ upon applying $B$ (Fig.~\ref{fig:3}c), with a slope indicating a Chern number of $C=-2$ as determined by the Streda formula ($\frac{dn}{dB} = C\frac{e}{h}$). These observations are all consistent with an incipient quantum anomalous Hall effect (QAHE) owing to a SVP state at $\nu=3$. Band structure modeling predicts a valley Chern number of $C_v=2$, consistent with our observation. The QAHE is not well developed, likely due to a combination of a small energy gap and substantial magnetic disorder in the sample~\cite{Grover2022}. These topological properties are reminiscent of those seen in t$1+2$ devices with similar twist angles~\cite{Polshyn2020}. 

The topological properties of t$2+3$ are more unusual. We again find a valley Chern number of $C_v=2$ in our band structure calculations and see clear signatures of an AHE, in this case below $\nu=1$ (shown in Fig.~\ref{fig:3}d for $\nu=0.75$). The magnitude of the AHE is small, but persists over a wide range of doping and vanishes very near $\nu=1$ (Fig.~\ref{fig:3}e). However, the corresponding Landau fan diagram exhibits an insulating state emerging from $\nu=1$ with zero slope up to high magnetic field (Fig.~\ref{fig:3}f), indicating that the $\nu=1$ symmetry-broken state is topologically trivial. Although the $C=0$ state may arise due to IVC ordering, the application of a large $B$ should favor a first-order phase transition to a SVP state. The absence of a first-order phase transition to a $C=2$ Chern insulator state is inconsistent with this scenario. An alternative possibility is that, upon opening a gap at $\nu=1$, interactions renormalize the Chern number of the filled band to $C=0$. In this scenario, the AHE can arise due to Berry curvature hot spots in the reconstructed bands formed by spontaneous symmetry breaking, while the total Berry curvature of the symmetry-broken band integrates to zero. Further theoretical and experimental work is needed to better resolve the nature of this state.

\medskip\noindent\textbf{Discussion and outlook}

Taken collectively, our measurements establish a new family of moir\'e graphene structures composed of Bernal-stacked graphene thin-film constituents with a small interfacial twist. By extending prior studies of twisted monolayer-bilayer and twisted double-bilayer graphene to thicker t$M+N$ variants, we discover a number of striking commonalities in both the single-particle and correlated states across this entire family. 

The localization of low-energy states to the layers at and just below the moir\'e interface, which underlies this phenomenon, may also generalize to moir\'e systems built from other vdW materials. For instance, a promising future direction would be to extend studies of twisted homobilayer transition metal dichalcogenides to similar multilayer constructions. This could, for example, help to establish new control knobs over the recently discovered fractional quantum anomalous Hall states in twisted bilayer MoTe$_2$~\cite{Cai2023,Park2023, Zeng2023}. Accurately modeling these structures will require the development of new theoretical analyses beyond those considered here, including the effects of electrostatic screening of $D$ and crystal fields at the twisted interface \cite{Rickhaus2019}. In parallel, further experimental studies into the largely unexplored physics of twisted monolayer-trilayer and bilayer-trilayer graphenes are very likely to reveal exciting new topological states. Our results thus establish a way to greatly expand the palette of topological flat bands available for study.

\section*{Methods}

\textbf{Device fabrication.} To fabricate t$M+N$ structures, we first optically identified an exfoliated graphene flake with a step, such that one portion of the flake is $M$ layers thick whereas another portion is $N$ layers. We next isolated regions of the $M$- and $N$-layer components using polymer-free anodic oxidation nanolithography~\cite{Li2018,Chen2019a,Saito2020}. We used standard dry transfer techniques with a polycarbonate (PC)/polydimethyl siloxane (PDMS) stamp~\cite{Wang2013} to stack isolated flakes with an interlayer twist by rotating the stage by an angle $\theta$ after the $M$-layer flake was picked up. These structures were encapsulated with hexagonal boron nitride flakes and graphite gates, and then transferred onto a Si/SiO$_2$ wafer. We used standard electron beam lithography and CHF$_3$/O$_2$ plasma etching to define vdW stacks into a Hall bar geometry and standard metal deposition techniques (Cr/Au)~\cite{Wang2013} to make electrical contact to the graphene multilayers. Optical images of all completed devices measured in this work are shown in Supplementary Fig.~1. 

We note that, in principle, there could be regions of the exfoliated graphene flakes with metastable stacking orders other than Bernal. However, these non-Bernal domains are known to relax to the ground-state Bernal stacking configuration during stacking unless great care is taken to isolate only non-Bernal domains and minimize strains during transfer~\cite{Zhou2021}. Since we do not take any such precautions, it is overwhelmingly likely that all non-Bernal domains relax to the Bernal configuration in our final samples. Another possible ambiguity in our sample construction is AB vs. BA stacking configurations between the twisted components. For example, t2+2 can host symmetry-broken states when twisted slightly away from $0^\circ$ (tAB+AB) or away from $60^\circ$ (tAB+BA)~\cite{He2023}. While we fabricated all of our devices by twisting slightly away from $0^\circ$, it is known that AB-BA stacking faults naturally exist in bilayer graphene flakes~\cite{Ju2015}, and these cannot be observed optically. It is therefore possible that t$M+N$ devices with $M\geq2$ and $N\geq2$ have unintentional domains of AB-BA at the twisted interface. These could, in principle, have different band structures and valley Chern numbers. Further work is needed to carefully distinguish these potential scenarios. 

\textbf{Transport measurements.} Transport measurements were carried out in various cryogen-free systems using a lock-in amplifier with frequencies between 13.33 and 17.77 Hz and a.c. bias between 1 and 10~nA. All data presented in the main text were acquired in Bluefors dilution refrigerators with nominal base temperatures between 10 and 30~mK, unless otherwise noted. Data reported at and above 1.5~K (e.g., Fig.~\ref{fig:3}d-e) were acquired in a Cryomagnetics variable temperature insert. 

The top and bottom gate voltages were used to independently control the carrier density, $n$, and perpendicular displacement field, $D$ according to the following relations: $n=(V_tC_t+V_bC_b)/e$ and $D=(V_tC_t-V_bC_b)/2\epsilon_0$, where $C_t$ and $C_b$
are the capacitance per unit area of the top and bottom dielectrics, respectively, $V_t$ and $V_t$ are the top and bottom gate voltages, respesctively, and $\epsilon_0$ is the vacuum permittivity. When specified, we perform field symmetrization (antisymmetrization) of $\rho_{xx}$ and $R_{xy}$ following $\rho_{xx} = [\rho_{xx}(B) + \rho_{xx}(-B)]/2$ and $R_{xy} = [R_{xy}(B) -R_{xy}(-B)]/2$.

We estimate the twist angle, $\theta$, between $M$- and $N$-layer flakes by fitting the sequences of quantum oscillations emerging from $\nu=0$ and $\nu=\pm4$ in Landau fan diagram measurements of $\rho_{xx}$ and $R_{xy}$ as a function of $n$ and magnetic field, $B$. From the Landau fan fit, we determine the superlattice density, $n_s$, and extract the twist angle using the relation $n_s=4\frac{2\theta^2}{\sqrt{3}a^2}$, where $a=0.246\ \rm{nm}$ is the graphene lattice constant.  

\textbf{Extraction of Fermi surface degeneracy from quantum oscillations.} 
The frequency of quantum oscillations, $f_\nu$, is extracted from low-field Landau fan measurements taken at constant $D$ (Extended Data Fig. \ref{fig:2-3_lowfield_QOs}). The frequency of oscillations, $f_{B}$, is first extracted from the Fourier transform (FFT) of each field sweep with respect to $1/B$. The frequency is then normalized by the total carrier density, $f_\nu=f_{B}/(\Phi_0 n)$. For the case of a singly connected Fermi surface at the Fermi level, the inverse quantity $f_\nu^{-1}$ represents the degeneracy of charge carriers~\cite{Zhou2021}. Extended Data Fig. \ref{fig:2-3_lowfield_QOs}b and c show the analysis outside of the correlated region of the t$2+3$ device, where there is a sharp peak in the FFT signal at $f_\nu^{-1}=4$, consistent with the four-fold degeneracy of graphene. Various line cuts are shown in the remainder of Extended Data Fig.~\ref{fig:2-3_lowfield_QOs}, and indicate quarter metal ($f_\nu^{-1}=1$) and half metal ($f_\nu^{-1}=2$) states formed over certain ranges of $n$ and $D$.

\textbf{Multiband transport in t$2+3$ graphene.} Regions of parameter space in the t$2+3$ device shown in Fig.~\ref{fig:3}a colored in white correspond to situations in which we are unable to unambiguously determine the degeneracy of the Fermi surface. Especially around $\nu=3$, this ambiguity likely arises due to multiple Fermi surface pockets coexisting within the moir\'e Brillouin zone. Extended Data Fig. \ref{fig:2-3-multiband transport} shows evidence for this in the form of curved trajectories of quantum Hall states seen in Landau fan diagrams. Such curved trajectories violate the Streda formula, which always predicts linear trajectories of topological gapped states, and generally arise due to the need to fill charge carriers simultaneously into two separate bands that each have their own sequence of Landau levels~\cite{Shi2018}. 

\textbf{Determination of isospin polarization at integer band fillings.} We employ a combination of out-of-plane and in-plane magnetic field measurements of $\rho_{xx}$ and $R_{xy}$ in order to infer the isospin polarization of the correlated states seen at $\nu=1$, $2$, and $3$. In-plane magnetic field couples primarily to the spin degree of freedom in graphene owing to its very weak spin-orbit coupling strength (although there can be orbital contributions in multilayer graphene samples~\cite{Ledwith2021}). Previous measurements of the correlated insulator at $\nu=2$ in t$1+2$ and t$2+2$ showed that the energy gap, as extracted from thermal activation measurements, grows with in-plane field~~\cite{Liu2020,Chen2021,Shen2020,Cao2020}. This behavior is consistent with spin-polarization, as the in-plane field adds a Zeeman contribution to the energy gap. We have performed similar measurements in our t$1+3$ ($\theta=1.29^{\circ}$) and t$2+3$ ($\theta=1.50^{\circ}$) devices. Extended Data Figs.~\ref{fig:spin-polarization}a-b show measurements of $\rho_{xx}$ as a function of $\nu$ in the moir\'e conduction band at various values of the in-plane magnetic field, $B_{||}$. In both devices, $\rho_{xx}$ is highly sensitive to $B_{||}$ very near $\nu=2$, exhibiting enhanced resistance for larger in-plane fields. These observations are consistent with spin-polarized correlated insulators at $\nu=2$ in t$1+3$ and t$2+3$. For the t$1+3$ device, we further performed temperature-dependent measurements at several values of $B_{||}$. In the Arrhenius plot shown in Extended Data Fig.~\ref{fig:spin-polarization}c, we can extract the gap size of the $\nu=2$ insulator in the thermally activated regime following $\rho_{xx}^{\nu=2} \propto e^{\Delta^{\nu=2}/2k_BT}$, where $\Delta^{\nu=2}$ is the gap size and $k_B$ is the Boltzmann constant. The results are shown in Extended Data Fig.~\ref{fig:spin-polarization}d. The measured gap size grows monotonically as a function of $B_{||}$. By fitting with a line, assuming $\Delta^{\nu=2}(B)= g\mu_B B + \Delta^{\nu=2}(0)$, we find $g\approx 2$, consistent with a spin-polarized insulating state. 

At $\nu=1$ and $3$, calculations typically find the most competitive ground states to be either IVC states or SVP states~\cite{Ledwith2021}. When $C_v$ is non-zero, these states are distinguished by the Chern number of the symmetry-broken state, which is $0$ for the IVC and non-zero for the SVP. A well-developed IVC state with non-zero $C_v$ would be a trivial insulator at integer band filling, whereas an SVP would exhibit the QAHE. Although these are in principle straightforward to distinguish, there can be various complicating factors in experiments. One example is twist-angle inhomogeneity in the sample, which can greatly obscure the QAHE. Another is that the correlated state may not be fully gapped at zero magnetic field, preventing a straightforward determination of the topology of the state. There are also exotic forms of IVC ordering that break time-reversal symmetry, leading to a metallic AHE~\cite{He2021tmbg}. Although band structure calculation of $C_v$ can provide guidance, the complexity of the calculation for thick t$M+N$ structures may result in incorrect predictions. Furthermore, interactions can potentially renormalize the Chern number of the symmetry-broken states at partial band filling. In general, we are not able to unambiguously determine the ground state ordering at $\nu=1$ and $3$ in our devices, except for the select cases shown in Fig.~\ref{fig:4} in the main text.

\textbf{Band structure calculations.} 
We utilize a generalized Bistrizter-MacDonald Hamiltonian for the single particle band structure calculations. The effective Hamiltonian can be written as 
\begin{equation*}\label{eq_Ham_all}
H=
\left(
\begin{array}{cc}
H_{M} & H_{\rm int}^{\dagger} \\
H_{\rm int} & H_{N} \\
\end{array}
\right),
\end{equation*}
where $H_M$ and $H_N$ are Hamiltonians for the $M$- and $N$-layer graphene, respectively, and $H_{\rm int}$ captures the interlayer coupling of the twisted \moire interface. The multilayer graphene Hamiltonians are given by
\begin{equation*}
H_{M}=
\left(
\begin{array}{ccccc}
H_{1} - \Delta_1     &        \Gamma   &   \Tilde{\Gamma}                        &  0  & \\
 \Gamma^{\dagger}   & H_{1} - \Delta_2 &   \Gamma^{\dagger} & \Tilde{\Gamma}' & \\
 \Tilde{\Gamma}  &  \Gamma  &    H_{1}- \Delta_3  & \Gamma & \\
 0 &\Tilde{\Gamma}' &    \Gamma^{\dagger}  &   H_1 -\Delta_4  & \\
 & &  & &   \ddots\\
\end{array}
\right),
\end{equation*}
with
\begin{equation*}
H_{1}=
\left(
\begin{array}{cc}
               0                             & \frac{\sqrt{3}}{2}\gamma_0 (k_x - i k_y) \\
\frac{\sqrt{3}}{2}\gamma_0 (k_x + i k_y) &                    0              \\
\end{array}
\right),
\end{equation*}
\begin{equation*}
\Gamma=
\left(
\begin{array}{cc}
-\frac{\sqrt{3}}{2}\gamma_4 (k_x - i k_y)      & -\frac{\sqrt{3}}{2}\gamma_3 (k_x + i k_y) \\
                     \gamma_1                  & -\frac{\sqrt{3}}{2}\gamma_4 (k_x - i k_y) \\
\end{array}
\right),
\end{equation*}

\begin{equation*} 
\Tilde{\Gamma}=
\left(
\begin{array}{cc}
\frac{1}{2}\gamma_2      & 0 \\
0  & \frac{1}{2}\gamma_5 \\
\end{array}
\right)
,\ \Tilde{\Gamma}'=
\left(
\begin{array}{cc}
\frac{1}{2}\gamma_5      & 0 \\
0  & \frac{1}{2}\gamma_2 \\
\end{array}
\right),
\end{equation*}
\begin{equation*} 
\rm{and}\ \Delta_i=
\left(
\begin{array}{cc}
\delta_i      & 0 \\
0  & \delta_i \\
\end{array}
\right)
\end{equation*}
The Hamiltonian is appropriately truncated according to how many layers there are. We use hopping parameters ($\gamma_0, \gamma_1, \gamma_2, \gamma_3, \gamma_4, \gamma_5) = (2610, 361, -20, -283, -140, 20)\ \rm{meV}$. The parameter $\delta_i$ captures the effect of a potential difference across the layers. For simplicity, we assume that the potential drops uniformly across the structure with a total magnitude given by $\delta=\sum_{i=1}^{M+N}|\delta_i|$. For example, the potentials for the t$1+2$ system would be $\delta_1=\delta/2, \delta_2 = 0,\ \rm{and}\ \delta_3=-\delta/2$. By this definition, $\delta>0$ corresponds to an experimentally applied $D>0$ that points from the top to the bottom, or thin to thick layer as shown in Fig. \ref{fig:1}. In the continuum approximation, the $M$- and $N$-layer systems are coupled when the Bloch wave vectors differ by $\Vec{q}_j$, where $\vec{q}_0=(0,0),\  \vec{q}_1=1/L_M\left(-\frac{2\pi}{\sqrt{3}}, -2\pi\right)$,  $\vec{q}_2=1/L_M\left(\frac{2\pi}{\sqrt{3}}, -2\pi\right)$, and $L_M=a/\theta$ is the \moire wavelength. The interlayer Hamiltonian is then given by
\begin{equation*}
H_{\rm int}=\sum_{j=0}^2 t_M
\left(
\begin{array}{cc}
\alpha      & \exp{\left(-i\frac{2\pi j}{3}\right)} \\
 \exp{\left(i\frac{2\pi j}{3}\right)}    & \alpha\\
\end{array}
\right),
\end{equation*}
where $t_M=110\ \rm{meV}$ and $\alpha=0.5$. 

Example band structures calculated for the t$1+3$ system are shown in Extended Data~\ref{fig:ED_theory_summary}a-c. Across many t$M+N$ constructions, we generally find overlap between the \moire conduction and valence bands with $\delta=0$. Further, the \moire valence band tends to be more dispersive than the \moire conduction band, consistent with our experimental observations that correlated states primarily occur on the electron-doped side. We performed a series of calculations for a range of $\theta$ and $\Delta$ for all t$M+N$ layer combinations up to t$3+6$, resulting in over 5000 individual band structures. To inform our experimental search, we quantify how isolated and flat the \moire conduction band is for each set of parameters by defining
\begin{equation*}
    \phi = \xi \frac{|\Delta_{+4}| |\Delta_{0}|}{\delta E},
\end{equation*}
where $\Delta_{+4}$ ($\Delta_{0}$) is the energy difference between the top (bottom) of the \moire conduction band and the remote conduction band (\moire valence band), and $\delta E$ is the bandwidth of the \moire conduction band (Extended Data~\ref{fig:ED_theory_summary}c). 
We define $\xi=+1$ when $\Delta_{+4}$ and $\Delta_{0}$ are both positive, and $\xi=-1$ otherwise. 

With this definition, $\phi$ becomes more positive as the moir\'e conduction band becomes more flat and isolated. When $\phi$ is negative, the \moire conduction band overlaps with the \moire valence band and/or the remote conduction band. Extended Data Fig.~\ref{fig:ED_theory_summary}d-g shows $\delta E$, $\Delta_{+4}$, $\Delta_0$, and $\phi$ as function of $\theta$ and $\delta$ for the t$1+3$ system, in which we find the optimal angle condition (defined as when $\phi$ achieves its largest positive value, shown in purple) to be at $\theta\approx1.30^{\circ}$ and $\delta\approx+75\ \rm{meV}$. Detailed results for each layer combination are shown in Supplementary Figs.~5-18. We summarize the results for all layer combinations in Extended Data Fig.~\ref{fig:ED_topology_summary}, where each data point is color-coded according to the maximum value of $\phi$ that is obtained for each system. Systems up to t$2+3$ have appreciably flat and isolated \moire conduction bands, but for t$1+4$, t$2+4$, and thicker, $\phi$ is very small or negative. For each layer combination in which we find that a flat and isolated \moire conduction band is predicted (i.e., $\phi_{max}>0$), we further calculate the valley Chern number of the \moire conduction band at the optimal parameter condition (i.e., for the values of $\theta$ and $\delta$ where $\phi=\phi_{\rm max}$). We find that all systems have non-zero $C_v$, reaching as large as $C_v=3$ for thicker layer combinations.  

Finally, we consider the effect of changing the Hamiltonian parameters to potentially explain the unexpected insulating state at $\nu=0$ and $D\approx0$ in t$2+3$ system (Fig.~\ref{fig:3}d-e). Within reasonable values of the tight binding parameters ($\gamma_i$), we find that there is always band overlap for twist angles between $1.33^{\circ}$ and $\theta=1.72^{\circ}$. The only parameter that has a significant effect at $\nu=0$ and $D=0$ is the strength of the \moire coupling, $t_M$. In Supplementary Fig. 19, we show band structure calculations varying $t_M$ around the nominal value of 110~meV. Increasing $t_M$ has the effect of flattening the bands further, but does not create a gap at charge neutrality. Reducing $t_M$ can create a small gap at charge neutrality, but in this case the \moire bands are much more dispersive and overlap with the remote bands. Both scenarios are inconsistent with the observations in our experiment. Taken together with the temperature dependence measurements shown in Fig.~\ref{fig:3}e, we conclude that the insulating state at $\nu=0$ and $D\approx0$ is most likely to be a correlated state.

\section*{Acknowledgements}
This work was supported by National Science Foundation (NSF) CAREER award no. DMR-2041972 and NSF MRSEC 1719797 at UW, and by NSERC, CFI, CIFAR and the Quantum Matter Institute at UBC. The development of twisted graphene samples is partially supported by the Department of Energy, Basic Energy Science Programs under award DE-SC0023062. XX. and M.Y. acknowledge support from the State of Washington-funded Clean Energy Institute. D.W. was supported by an appointment to the Intelligence Community Postdoctoral Research Fellowship Program at University of Washington administered by Oak Ridge Institute for Science and Education through an interagency agreement between the US Department of Energy and the Office of the Director of National Intelligence. E.T. and E.A.-M. were supported by grant no. NSF GRFP DGE-2140004. K.W. and T.T. acknowledge support from the Elemental Strategy Initiative conducted by the MEXT, Japan (grant no. JPMXP0112101001) and JSPS KAKENHI (grant nos. 19H05790, 20H00354 and 21H05233). Y.-H.Z. was supported by the National Science Foundation under Grant No. DMR-2237031. This work made use of shared fabrication facilities provided by NSF MRSEC 1719797. This research acknowledges usage of the millikelvin optoelectronic quantum material laboratory supported by the M.J. Murdock Charitable Trust.

\textbf{Author contributions.} D.W., E.T., A.O., E.A.-M., M.H., and K.H. fabricated the devices. D.W., R.S., E.T., and M.H. performed the measurements and analyzed data. Y.Z. wrote the code to calculate the continuum model band structures. D.W. performed the continuum model calculations. K.W. and T.T. grew the BN crystals. X.X., J.F., and M.Y. supervised device fabrication, measurement, and data analysis. D.W., R.S., E.T., J.F., and M.Y. wrote the paper with input from all authors.

\section*{Competing interests}
The authors declare no competing interests.

\section*{Additional Information}
Correspondence and requests for materials should be addressed to Joshua Folk or Matthew Yankowitz.

\section*{Data Availability}
Source data are available for this paper. All other data that support the findings of this study are available from the corresponding author upon request.

\bibliographystyle{naturemag}
\bibliography{references}

\clearpage

\renewcommand{\figurename}{Extended Data Fig.}
\renewcommand{\thesubsection}{S\arabic{subsection}}
\setcounter{secnumdepth}{2}
\setcounter{figure}{0} 
\setcounter{equation}{0}

\onecolumngrid

\section*{Extended Data}

\begin{figure*}[h]
\includegraphics[width=0.75\textwidth]{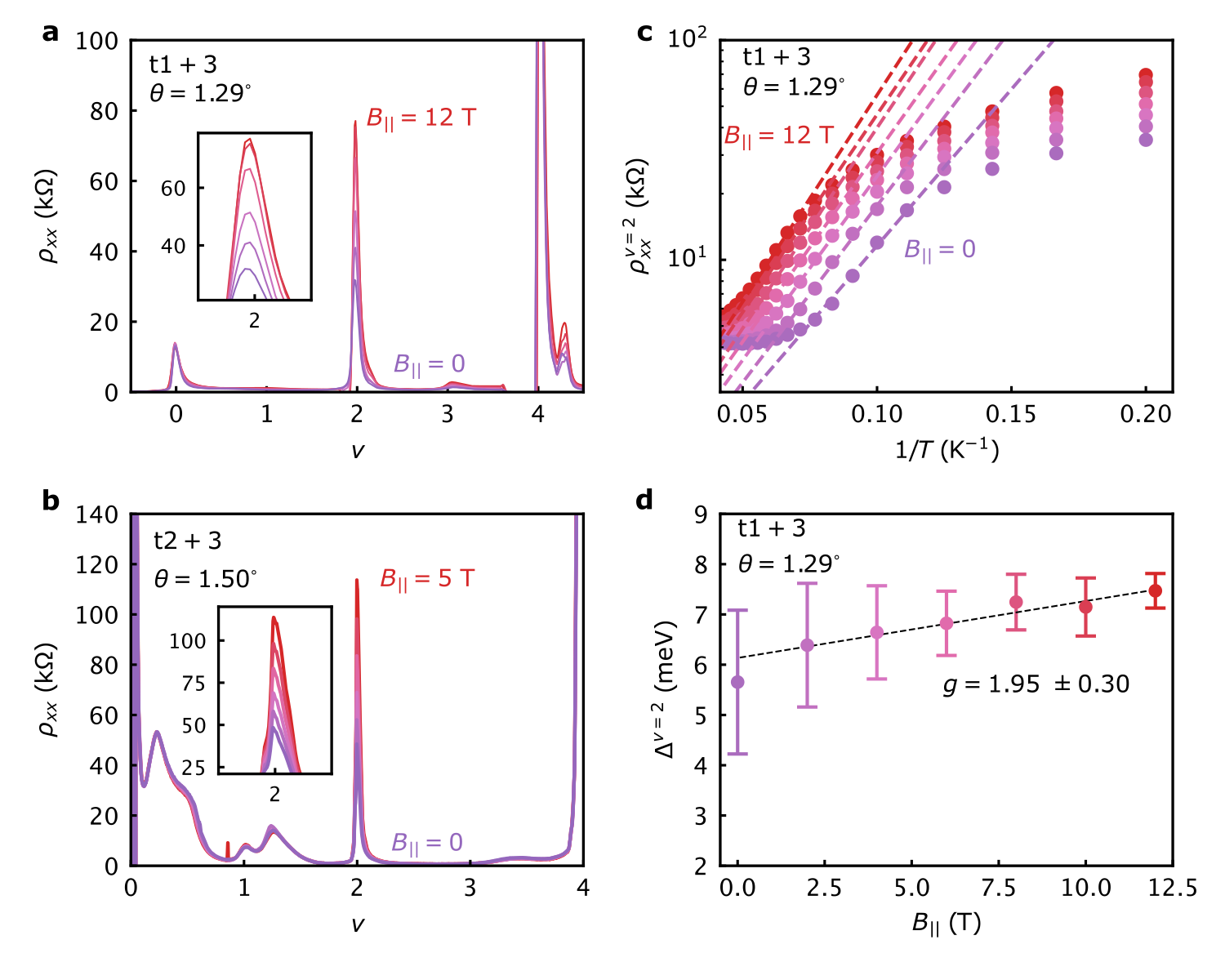} 
\caption{\textbf{Evidence for spin-polarization at $\bm{\nu=2}$.}
\textbf{a}, Measurement of $\rho_{xx}$ versus $\nu$ for the t$1+3$ sample with $\theta = 1.29^{\circ}$ acquired at different values of $B_{||}$. The measurement is acquired at variable $D$ in order to cut through the correlated states at each integer $\nu$. The inset shows a zoom-in around $\nu=2$, showing an increase in the resistance of the correlated insulator with $B_{||}$. The measurement was performed at $T=1.7$~K. 
\textbf{b}, Similar measurement for the t$2+3$ device with $\theta = 1.50^{\circ}$, acquired at $D=0.531$~V/nm and $T=4$~K. 
\textbf{c}, $\rho_{xx}$ acquired at $\nu=2$ in the t$1+3$ device as a function of temperature, shown at different values of $B_{||}$ in increments of 2~T. The dashed lines show the linear fits in the thermally activated regime used to extract the gap size of the correlated insulator (see Methods). 
\textbf{d}, Extracted gap sizes determined by the fits shown in \textbf{c}. Error bars are standard deviations from the fits. The dashed black line shows the best linear fit to the experimentally determined gap sizes. The slope yields a $g$-factor consistent with $g=2$ (see Methods), consistent with a spin-polarized insulator. 
} 
\label{fig:spin-polarization}
\end{figure*}

\begin{figure*}[h]
\includegraphics[width=\textwidth]{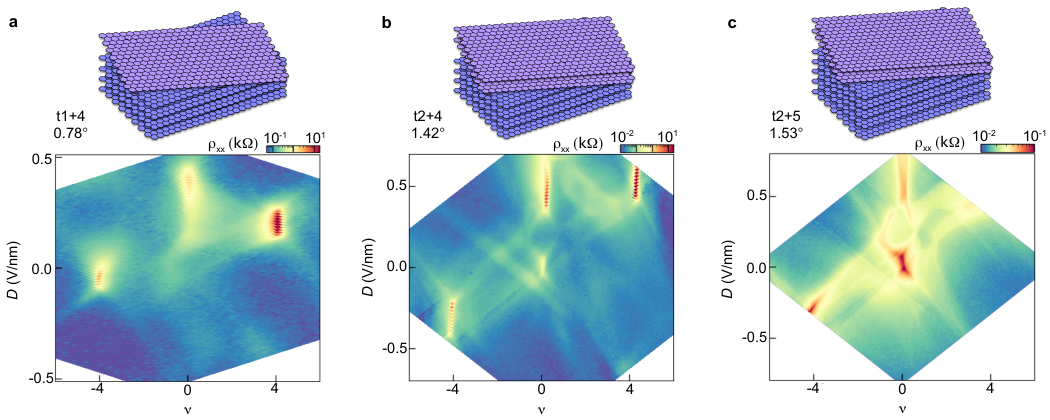} 
\caption{\textbf{Additional t$\bm{M+N}$ devices.}
\textbf{a}, $\rho_{xx}$ map of a t$1+4$ device with $\theta=0.78^{\circ}$ acquired at $B=0$~T. High-resistance states emerge over a small range of $D$ at $\nu=0$ and $\nu=\pm4$, indicating the emergence of isolated moir\'e bands. 
\textbf{b}, Similar measurement for a t$2+4$ device with $\theta=1.42^{\circ}$.
\textbf{c}, Similar measurement for a t$2+5$ device with $\theta=1.53^{\circ}$. 
Schematic above each map shows cartoon representations of the given twisted $M$- and $N$-layer combination. Measurements for \textbf{a-b} were acquired at $T=1.5$~K. Measurements for \textbf{c} were acquired at $T=100$~mK. 
} 
\label{fig:additional_devices}
\end{figure*}

\begin{figure*}[h]
\includegraphics[width=\textwidth]{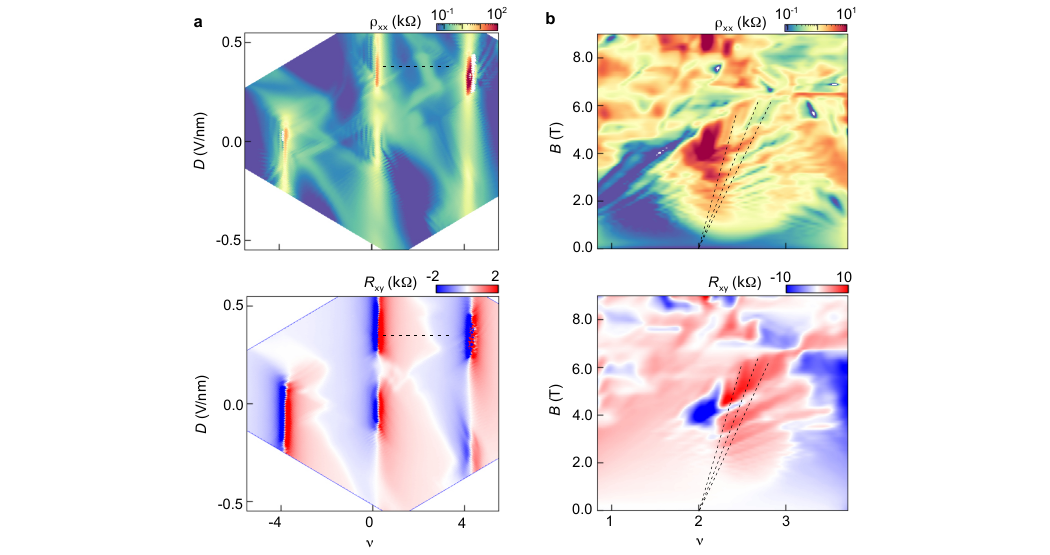} 
\caption{\textbf{High-field correlated state in a t$\bm{2+4}$ device.}
\textbf{a}, $\rho_{xx}$ (top) and $R_{xy}$ (bottom) maps for the t$2+4$ device with $\theta=1.15^{\circ}$ (the same as shown in Fig.~\ref{fig:1}h of the main text). The longitudinal (Hall) maps are symmetrized (anti-symmetrized)
at $B=0.5$~T. 
\textbf{b}, Landau fan diagrams acquired at $D=0.35$~V/nm over the range of doping range indicated by dashed line in \textbf{a}. A high-resistance state accompanied by an abrupt sign reversal in $R_{xy}$ emerges at $\nu=2$ for $B\approx4$~T. There are additionally quantum oscillations associated with this state that project to $\nu=2$ at $B=0$. Together, these observations indicate the emergence of a symmetry-broken state at $\nu=2$ in a magnetic field. This state is destroyed at higher field by competition with other quantum Hall states. All measurements were acquired at $T=100$~mK. 
}
\label{fig:2-4_LF}
\end{figure*}

\begin{figure*}[t]
\includegraphics[width=\textwidth]{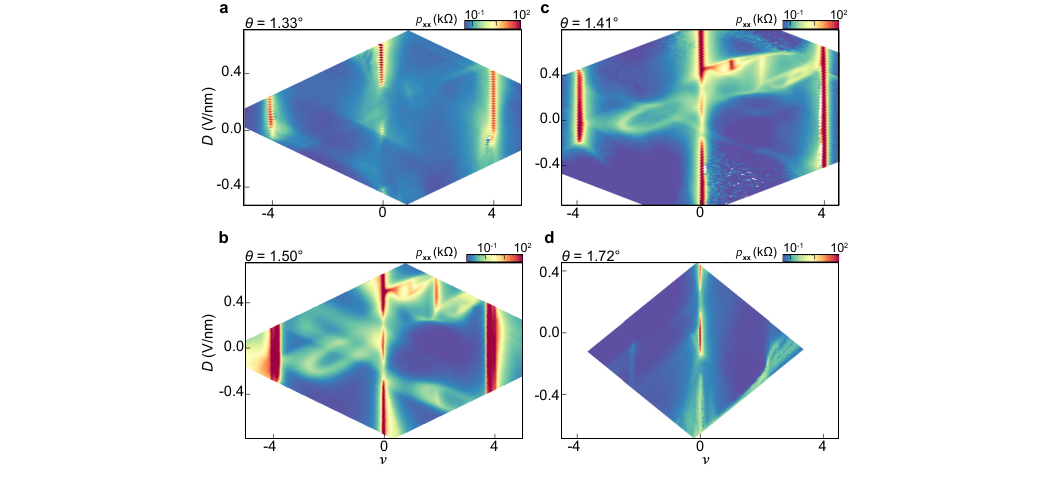} 
\caption{\textbf{Twist angle dependence of t2+3.} 
$\rho_{xx}$ map as a function of $\nu$ and $D$ for four 2+3 devices with different twist angles: \textbf{a}, $\theta=1.33^{\circ}$ \textbf{b}, $1.41^{\circ}$ \textbf{c}, $1.56^{\circ}$, and \textbf{d} $1.72^{\circ}$. Data shown in \textbf{a}, \textbf{b} and \textbf{c} feature insulating states at $\nu=0$ and $\nu=\pm4$. Only devices shown in \textbf{b} and \textbf{c} exhibit correlated insulating states at positive $D$ for partial filling of the moir\'e conduction band. In addition, \textbf{c} further shows a correlated insulating state at negative $D$. All data taken at $T=1.5\ \rm{K}$.
}
\label{fig:2-3_twist_dep}
\end{figure*}

\begin{figure*}[h]
\includegraphics[width=\textwidth]{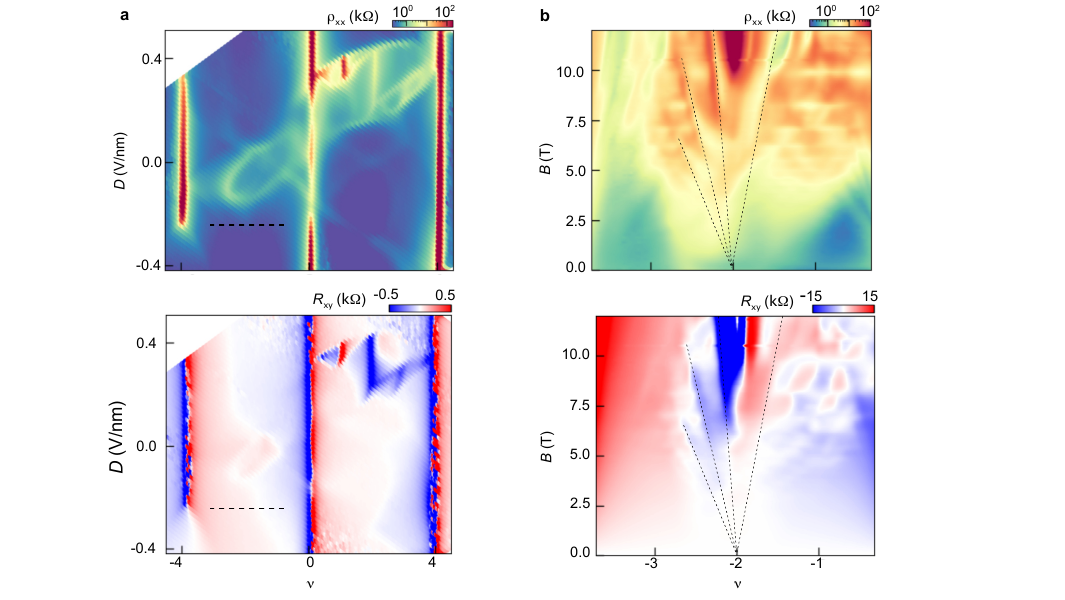} 
\caption{\textbf{Correlated states at high field in a t2+3 device with $\bm{\theta=1.41^{\circ}}$.} 
\textbf{a}, $\rho_{xx}$ (top) and $R_{xy}$ (bottom) maps. The longitudinal (Hall) maps are symmetrized (anti-symmetrized) at $B=0.1$~T 
\textbf{b}, Landau fans acquired at $D=-0.25$ V/nm over the doping range indicated by dashed line in \textbf{a}. The data shows clear quantum oscillations projecting to $\nu=-2$ at $B=0$, associated with a high-resistance state accompanied by an abrupt sign reversal in $R_{xy}$ above $B\approx10$~T. Together, these observations indicate the emergence of a symmetry-broken state at $\nu=2$ in a magnetic field. All measurements were acquired at $T=1.5$~K. 
}
\label{fig:2-3_high_field_correlations}
\end{figure*}

\begin{figure*}[h]
\includegraphics[width=\textwidth]{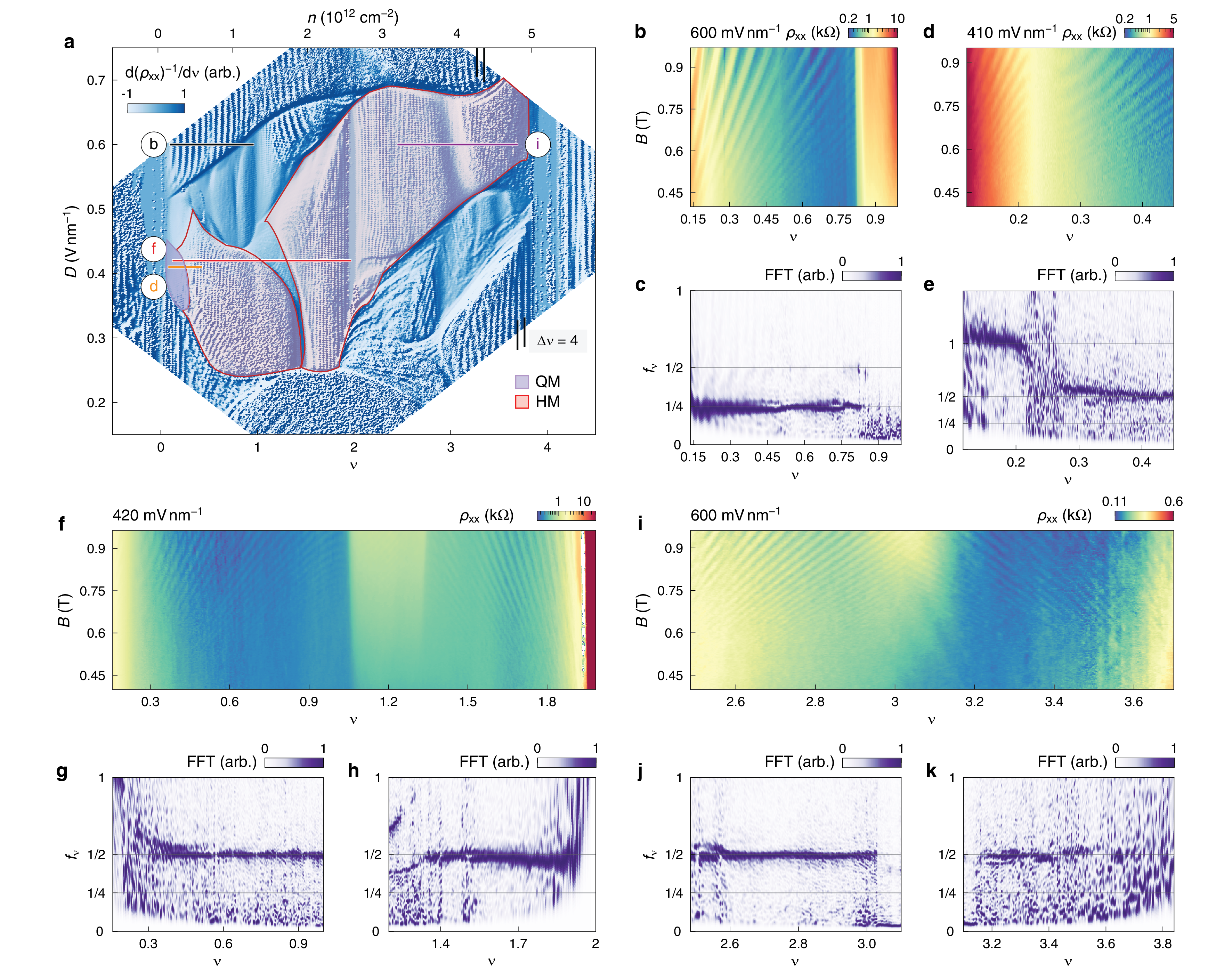} 
\caption{\textbf{Low-field quantum oscillations in the t2+3 device with $\bm{\theta=1.50^{\circ}}$.} 
\textbf{a}, Numerical derivative of the inverse of $\rho_{xx}$ with respect to $\nu$, chosen to highlight the quantum oscillations. Gate voltages where two/one fold-degenerate oscillations ($\Delta \nu = 2$/$\Delta \nu = 1$) (shaded in red/purple) correspond to the half/quarter metal regions in the phase diagram shown in Fig.~\ref{fig:3}a of the main text (only a subset of all symmetry-broken regions are highlighted here for clarity). The raw data is the same as shown in Fig.~\ref{fig:2}d of the main text. Alphabetical labels b-i associated with lines of constant $D$ show where data in the panels with the same label were obtained. 
\textbf{b}, \textbf{c}, Landau fan diagram obtained at $D = 600$~mV/nm, and corresponding Fourier transform analysis. The dominant FFT peak occurs at $f_{\nu} = 1/4$, showing that the area of a cyclotron orbit is 1/4 of the total Luttinger volume, corresponding to a normal metal with four-fold spin and valley degeneracy. 
\textbf{d}, \textbf{e}, Landau fan obtained at $D = 410$ mV/nm, and corresponding FFT. 
\textbf{f}, Landau fan measured at $D = 420$ mV/nm. 
\textbf{g}, \textbf{h},  Corresponding FFTs of \textbf{f}. \textbf{i}, Landau fan measured at $D = 600$ mV/nm for $\nu > 2$. 
\textbf{j}, \textbf{k}, Corresponding FFTs of \textbf{i}. The spectra at each band filling have been normalized by the maximum FFT amplitude at that band filling.  In addition, when computing $f_{\nu}$ for quantum oscillations that originate from a finite band filling, the zero of carrier density was set to that band filling. In \textbf{h} and \textbf{j}, the zero was set to $\nu = 2$, and in \textbf{k}, the zero was set to $\nu = 4$. However, the horizontal axes display $\nu$ without any offsets for ease of comparison. With the exception of \textbf{a}, which was obtained at $T = 100$~mK, the remaining panels were obtained at the base mixing chamber temperature of $T = 10$~mK.
} 
\label{fig:2-3_lowfield_QOs}
\end{figure*}

\begin{figure*}[h]
\includegraphics[width=\textwidth]{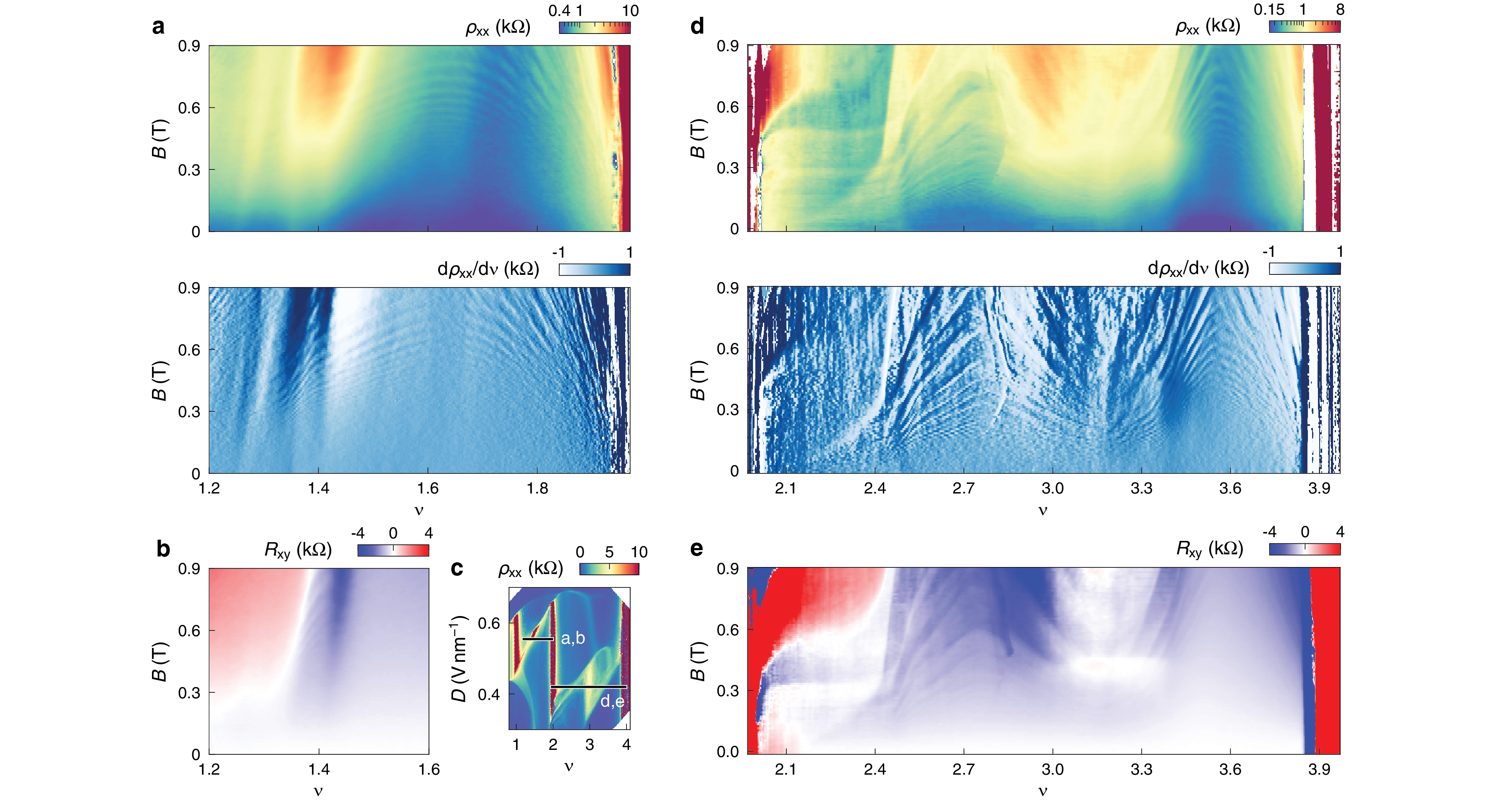} 
\caption{\textbf{Signatures of multiband transport in the t2+3 device with $\bm{\theta=1.50^{\circ}}$.} 
\textbf{a}, Longitudinal resistance and its numerical derivative obtained between $\nu = 1$ and $2$, highlighting the evolution of curved quantum Hall states developing at low-$B$. \textbf{b}, Hall resistance as a function of $B$. \textbf{c}, Zoomed-in view of the phase diagram shown in the main text for reference. Horizontal lines indicate the displacement fields $D = 0.555$ V/nm and $D = 0.450$ V/nm where panels (\textbf{a},\textbf{b}) and (\textbf{d},\textbf{e}) were respectively obtained. \textbf{d}, \textbf{e}, Longitudinal and Hall resistance obtained near $\nu = 3$. Measurements were obtained at the base mixing chamber temperature of $T = 10$~mK.
}
\label{fig:2-3-multiband transport}
\end{figure*}

\begin{figure*}[h]
\includegraphics[width=\textwidth]{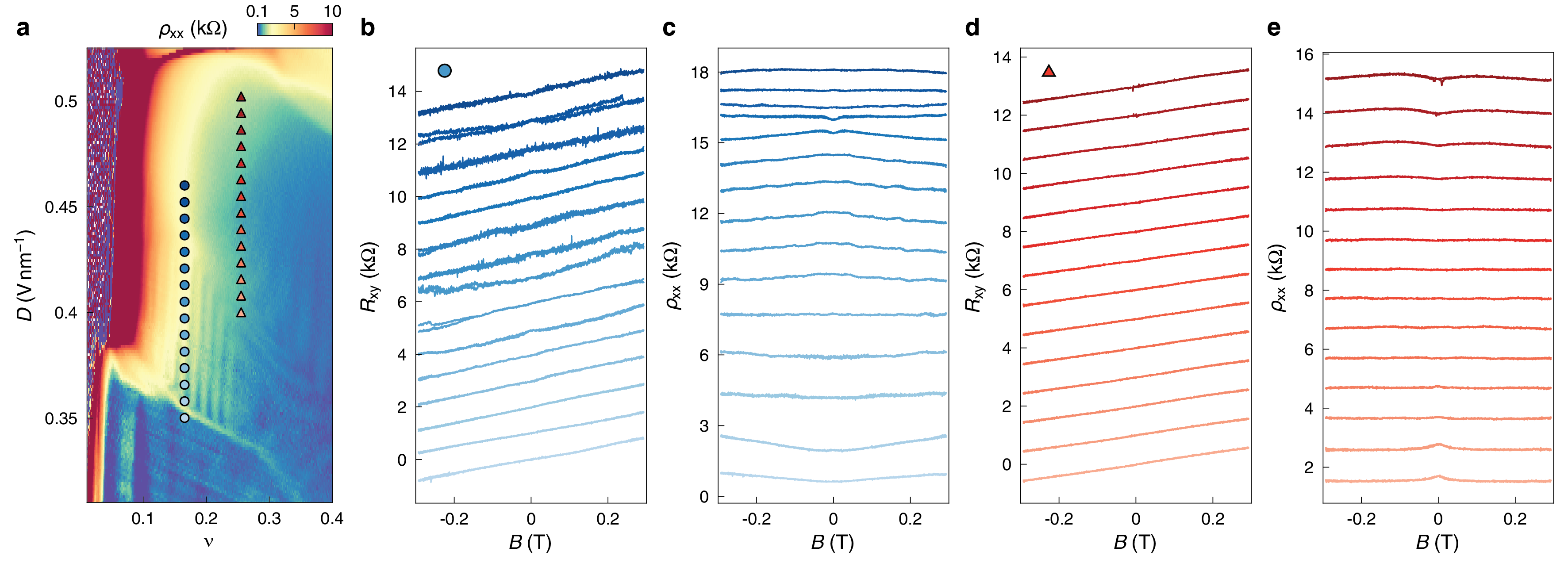} 
\caption{\textbf{Hall and longitudinal magnetoresistance at low density in the t2+3 device with $\bm{\theta=1.50^{\circ}}$.} \textbf{a}, High resolution $\rho_{xx}$ map reproduced from Fig.~\ref{fig:3}b, focusing on regions that show no quantum oscillations (near $\nu = 0.25$, $D$ between 0.45 and 0.5 V/nm) or fully degeneracy lifted quantum oscillations (near $\nu = 0.2$, $D$ between 0.35 and 0.45 V/nm). \textbf{b}, \textbf{c}, $\rho_{xy}$, $\rho_{xx}$ measured by sweeping $B$ back and forth, obtained at fixed $\nu = 0.17$ (circle markers in \textbf{a}). Traces at different $\nu$, $D$ settings have been offset by 1~k$\Omega$. \textbf{d}, \textbf{e}, Similar measurements to \textbf{b} and \textbf{c}, obtained at fixed $\nu = 0.26$ (triangle markers in \textbf{a}). Data at each location have been offset by 1k$\Omega$ for clarity. None of the measurements show evidence for the anomalous Hall effect, even in the quarter-metal region. Measurements were obtained at $T = 100$~mK.
} 
\label{fig:2-3_noAHE}
\end{figure*}

\begin{figure*}[h]
\includegraphics[width=0.86\textwidth]{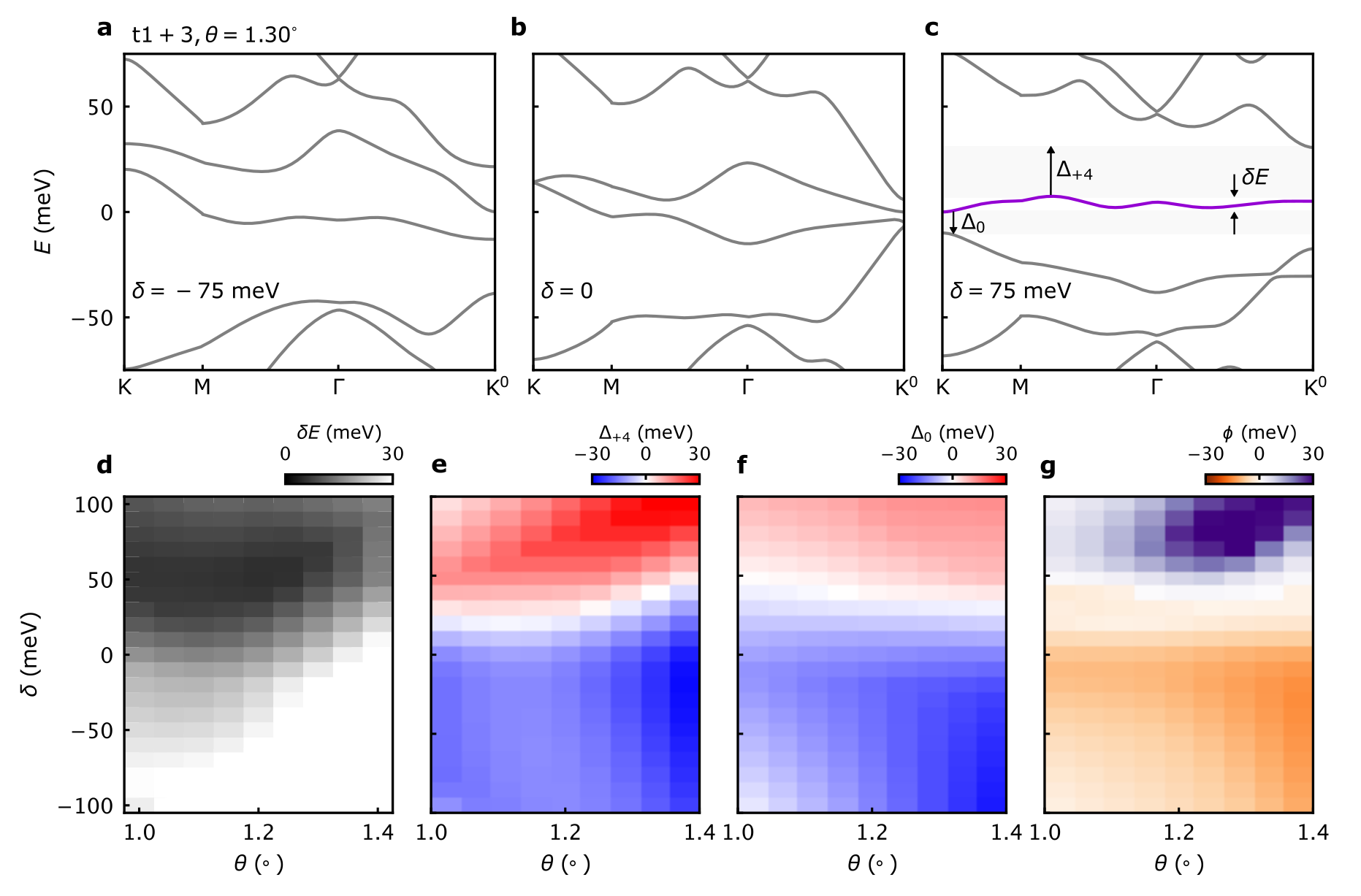} 
\caption{\textbf{Band structure calculations for t$\mathbf{1+3}$.}
\textbf{a-c}, Band structures for the t$1+3$ system at $1.30^\circ$, with interlayer potentials $\delta=-75, 0,$ and $75\ \rm{meV}$, respectively. Panel \textbf{c} shows the definitions of $\Delta_{0}$, $\Delta_{+4}$, and $\delta E$.
\textbf{d}, Bandwidth, $\delta E$, of the \moire conduction band calculated as a function of $\theta$ and $\delta$. 
\textbf{e}, Energy separation between the top of the \moire conduction band and the top of the remote conduction band, $\Delta_{+4}$. Positive (negative) values of $\Delta_{+4}$ indicate a band gap (band overlap). 
\textbf{f}, Similar plot, but for the energy separation between the \moire conduction and valence band, $\Delta_0$ (i.e., the gap at the charge neutrality point).
\textbf{g}, The isolated flat band parameter, $\phi$ (see Methods), as a function of twist angle and interlayer potential. Large positive values of $\phi$ (purple), corresponds to a flat and isolated \moire conduction band. Negative values of $\phi$ (orange) correspond to the \moire conduction band overlapping with other bands. The optimal twist angle is around $1.30^{\circ}$, corresponding to the band structures shown in panels \textbf{a}-\textbf{c}.
} 
\label{fig:ED_theory_summary}
\end{figure*}

\begin{figure*}[h]
\includegraphics[width=4.in]{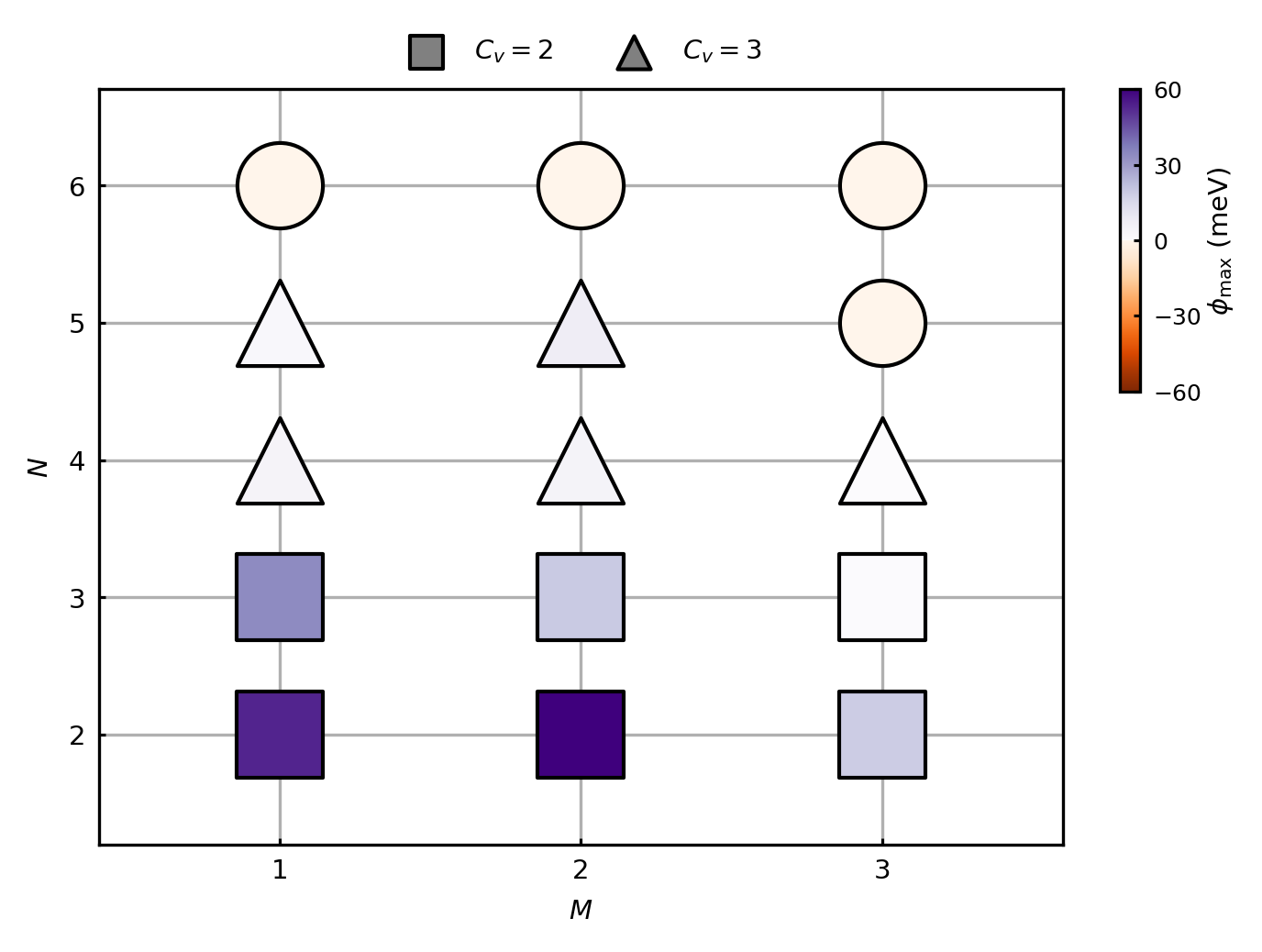} 
\caption{\textbf{Calculated band isolation, flatness, and topology of various t$\bm{M+N}$ structures.}
Summary of the band structure calculations for all t$M+N$ combinations up to t$3+6$. The color of each marker indicates the maximum value of $\phi$ obtained after searching over all combinations of $\theta$ and $\delta$. The valley Chern number of the \moire conduction band for each layer combination is also calculated for the optimal parameters of $\theta$ and $\delta$, i.e. where $\phi=\phi_{\rm max}$. The shape of the markers indicates the valley Chern number, where squares denote $C_v=2$ and triangles denote $C_v=3$. We do not calculate the valley Chern number for systems where the \moire conduction band cannot be isolated ($\phi_{\rm max}<0$), indicated by the circle markers. 
} 
\label{fig:ED_topology_summary}
\end{figure*}

\end{document}



\title{Supplementary Information: Topological flat bands in a family of multilayer graphene \moire lattices}

\author{Dacen Waters$^{1,2*}$}
\author{Ruiheng Su$^{3,4*}$}
\author{Ellis Thompson$^{1*}$}
\author{Anna Okounkova$^{1}$}
\author{Esmeralda Arreguin-Martinez$^{5}$}
\author{Minhao He$^{1, 6}$}
\author{Katherine Hinds$^{5}$}
\author{Kenji Watanabe$^{7}$} 
\author{Takashi Taniguchi$^{8}$} 
\author{Xiaodong Xu$^{1,5}$}
\author{Ya-Hui Zhang$^{9}$}
\author{Joshua Folk$^{3,4\dagger}$}
\author{Matthew Yankowitz$^{1,5\dagger}$}

\affiliation{$^{1}$Department of Physics, University of Washington, Seattle, Washington, 98195, USA}
\affiliation{$^{2}$ Intelligence Community Postdoctoral Research Fellowship Program, University of
Washington, Seattle, Washington, 98195, USA}
\affiliation{$^{3}$ Stewart Blusson Quantum Matter Institute, University of British Columbia, Vancouver, British Columbia, V6T 1Z1, Canada}
\affiliation{$^{4}$ Department of Physics and
Astronomy, University of British Columbia, Vancouver, British Columbia, V6T 1Z1, Canada}
\affiliation{$^{5}$Department of Materials Science and Engineering, University of Washington, Seattle, Washington, 98195, USA}
\affiliation{$^{6}$Department of Physics, Princeton University, Princeton, New Jersey, 08544, USA}
\affiliation{$^{7}$Research Center for Functional Materials,
National Institute for Materials Science, 1-1 Namiki, Tsukuba 305-0044, Japan}
\affiliation{$^{8}$International Center for Materials Nanoarchitectonics,
National Institute for Materials Science, 1-1 Namiki, Tsukuba 305-0044, Japan}
\affiliation{$^{9}$Department of Physics and Astronomy, Johns Hopkins University, Baltimore, Maryland, 21205, USA}
\affiliation{$^{*}$These authors contributed equally to this work.}
\affiliation{$^{\dagger}$ jfolk@physics.ubc.ca (J.F.); myank@uw.edu (M.Y.)}

\maketitle

\newpage

\begin{figure*}[h]
\includegraphics[width=\textwidth]{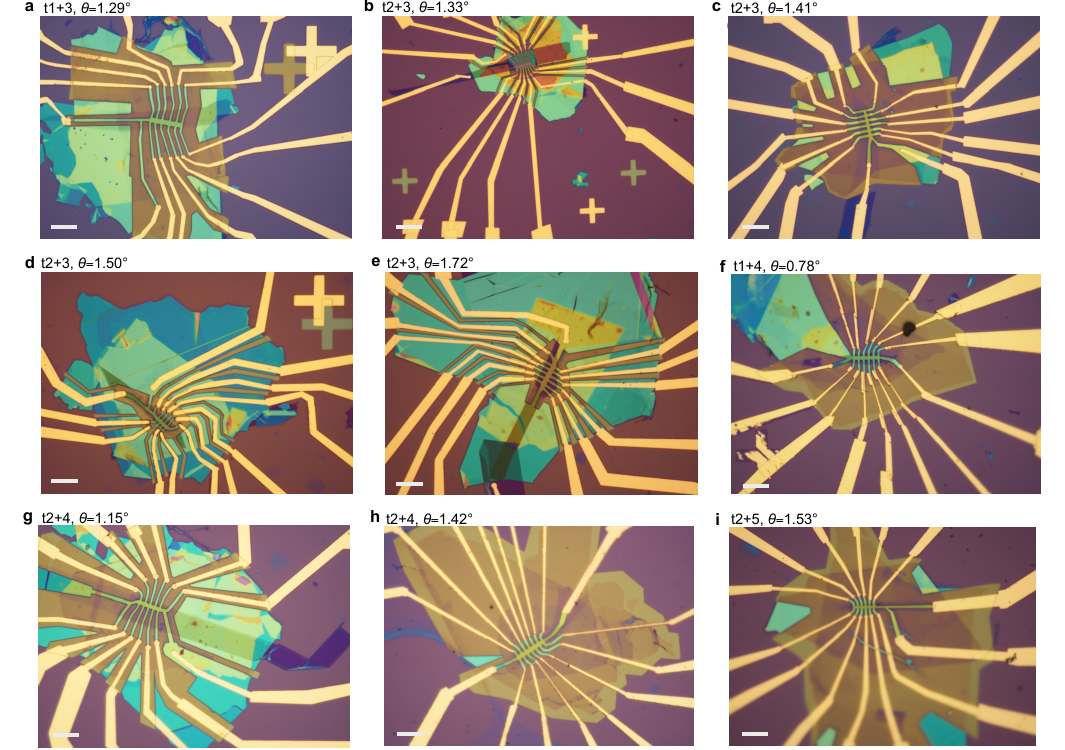} 
\caption{\textbf{Optical micrographs of the t$\bm{M+N}$ devices in this study.} 
\textbf{a-i,} All scale bars are 10 $\mu m$.
}
\label{fig:devices}
\end{figure*}

\newpage

\section*{Additional transport data}

\begin{figure*}[h]
\includegraphics[width=\textwidth]{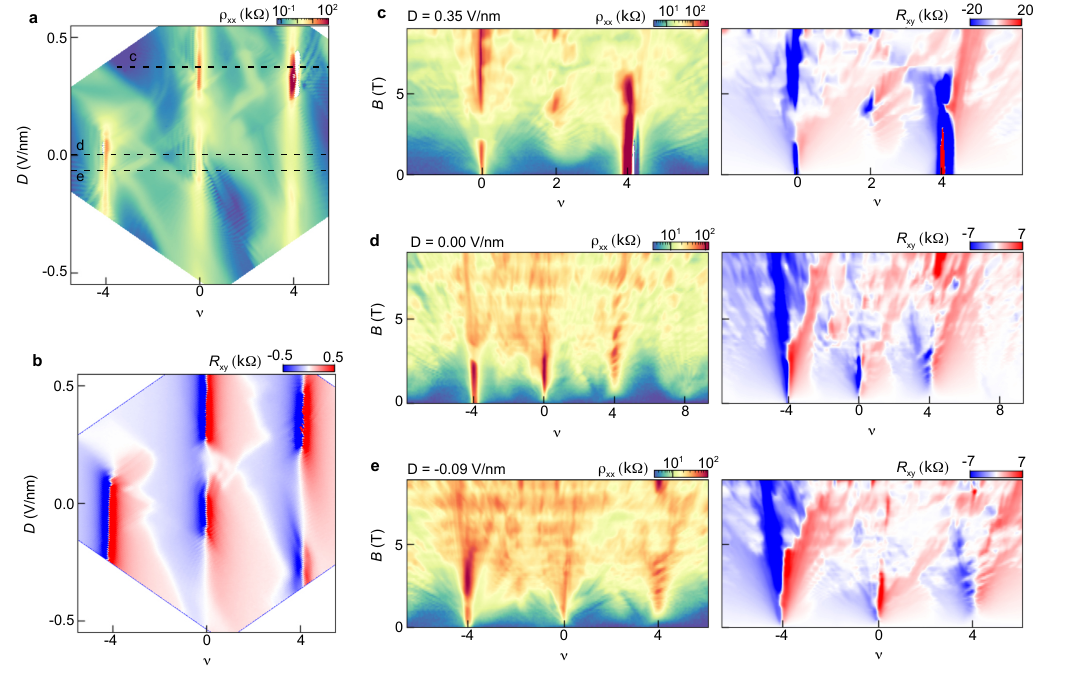} 
\caption{\textbf{Additional Data for t$\bm{2+4}$. } 
\textbf{a}, Longitudinal and \textbf{b,} Hall resistance maps in the t2+4 device $\theta=1.42^{\circ}$. The longitudinal (Hall) maps are symmetrized (anti-symmetrized) at $B=0.5$~T. 
\textbf{c-e}, Landau fan diagrams acquired at fixed displacement fields of \textbf{(c)} $D=0.35$~V/nm, \textbf{(d)} $D=0$, and \textbf{(e)} $D=-0.09$~V/nm. All measurements in figure were acquired at $T=100$~mK. 
}
\label{fig:SI_2p4}
\end{figure*}

\begin{figure*}[h]
\includegraphics[width=\textwidth]{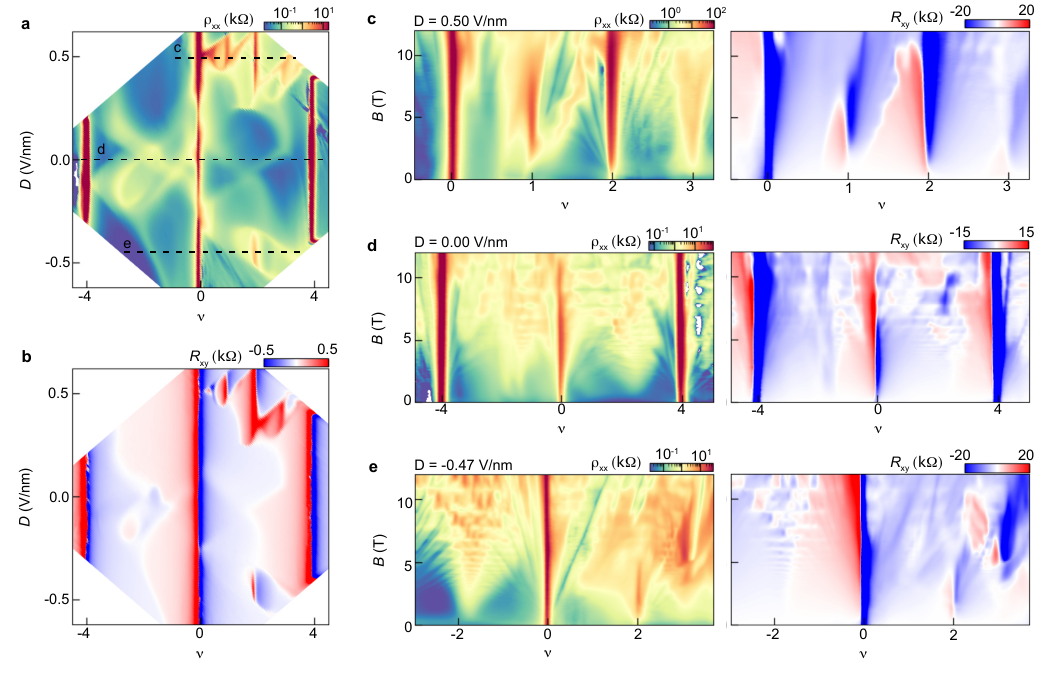} 
\caption{\textbf{Additional Data for t$\bm{2+3}$.}
\textbf{a}, Longitudinal and \textbf{b,} Hall resistance maps in the t2+3 device with $\theta=1.50^{\circ}$. The longitudinal (Hall) maps are symmetrized (anti-symmetrized)
at $B=0.1$~T. 
\textbf{c-e}, Landau fan diagrams acquired at fixed displacement fields of \textbf{(c)} $D=0.50$ V/nm, \textbf{d} $D=0$, and \textbf{e} $D=-0.47$ V/nm. All measurements in figure were acquired at $T=1.5$~K.} 
\label{fig:1-3_LF}
\end{figure*}

\begin{figure*}[h]
\includegraphics[width=\textwidth]{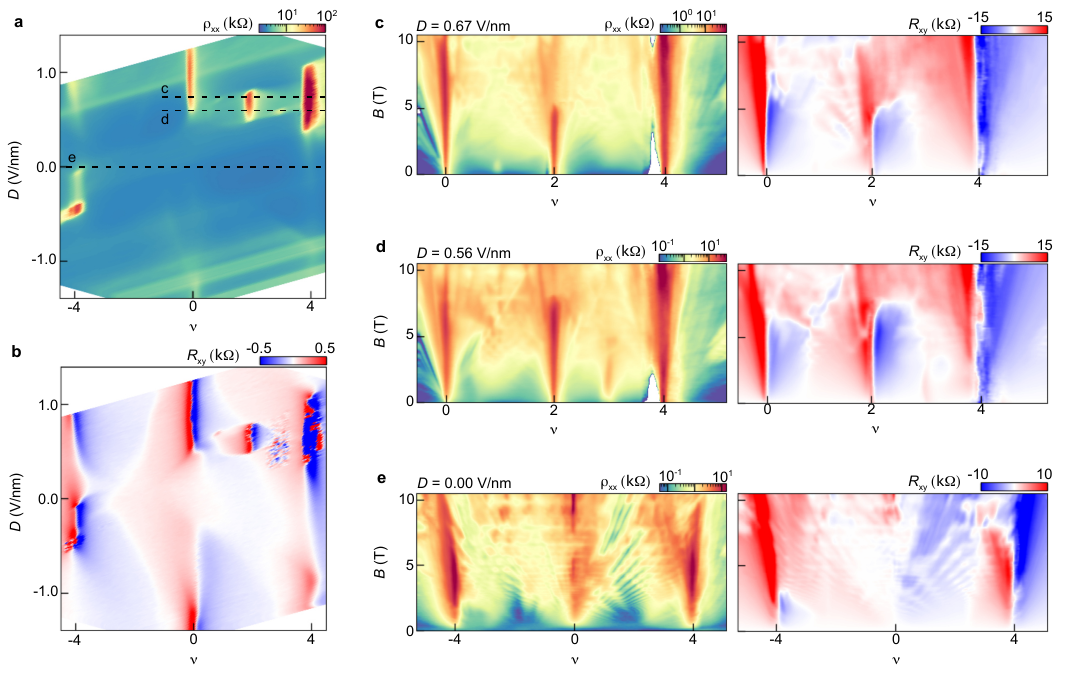} 
\caption{\textbf{Additional Data for t$\bm{1+3}$. } 
\textbf{a,} Longitudinal and \textbf{b,} Hall resistance maps in the t1+3 device. The longitudinal (Hall) maps are symmetrized (anti-symmetrized)
at $B=$ 0.5 T. \textbf{c-e,} Longitudinal and Hall resistances as a function of $B$ at fixed displacement field of \textbf{c} $D=0.67$ V/nm, \textbf{d} $D=0.56$ V/nm, \textbf{e} $D=-0.00$ V/nm. All measurements in figure were acquired at $T=$ 1.5 K.
}
\label{fig:SI_2p4}
\end{figure*}

\section*{Continuum model calculation results}

\begin{figure*}[h]
\includegraphics[width=0.9\textwidth]{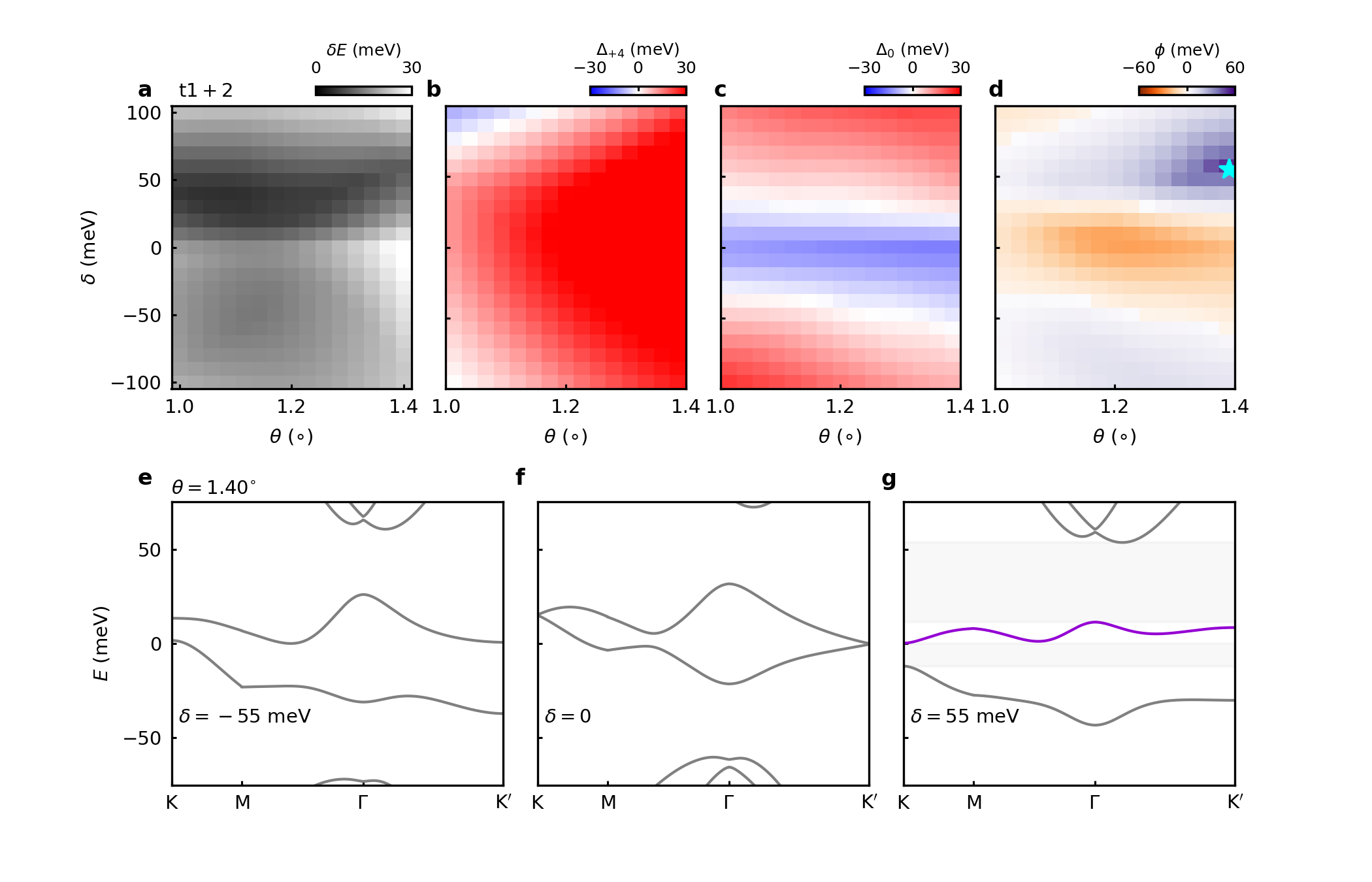} 
\caption{\textbf{Single particle calculations for t$\bm{1+2}$.} 
\textbf{a}, Bandwidth of the \moire conduction band, $\delta E$ as a function of twist angle, $\theta$, and interlayer potential, $\delta$. 
\textbf{b}, Energy difference between the top of the \moire conduction band and the remote conduction band, $\Delta_{+4}$. Positive values indicate a gap, negative values indicate band overlap.
\textbf{c}, Energy difference between the bottom of the \moire conduction band and the top of the \moire valence band, $\Delta_0$. 
\textbf{d}, Isolated flat band parameter, $\phi$, see Methods for definition. Larger positive values indicates more flat and isolated \moire conduction band. Cyan star denotes the optimal parameter condition for a flat and isolated \moire conduction band. 
\textbf{e}-\textbf{g}, Representative band structures at the optimal angle condition at the indicated values of $\delta$. In panel \textbf{g}, the \moire conduction band is highlighted in purple and shaded regions indicate gaps, i.e. it is isolated from other bands. 
}
\label{fig:phase_t12}
\end{figure*}

\begin{figure*}[h]
\includegraphics[width=0.9\textwidth]{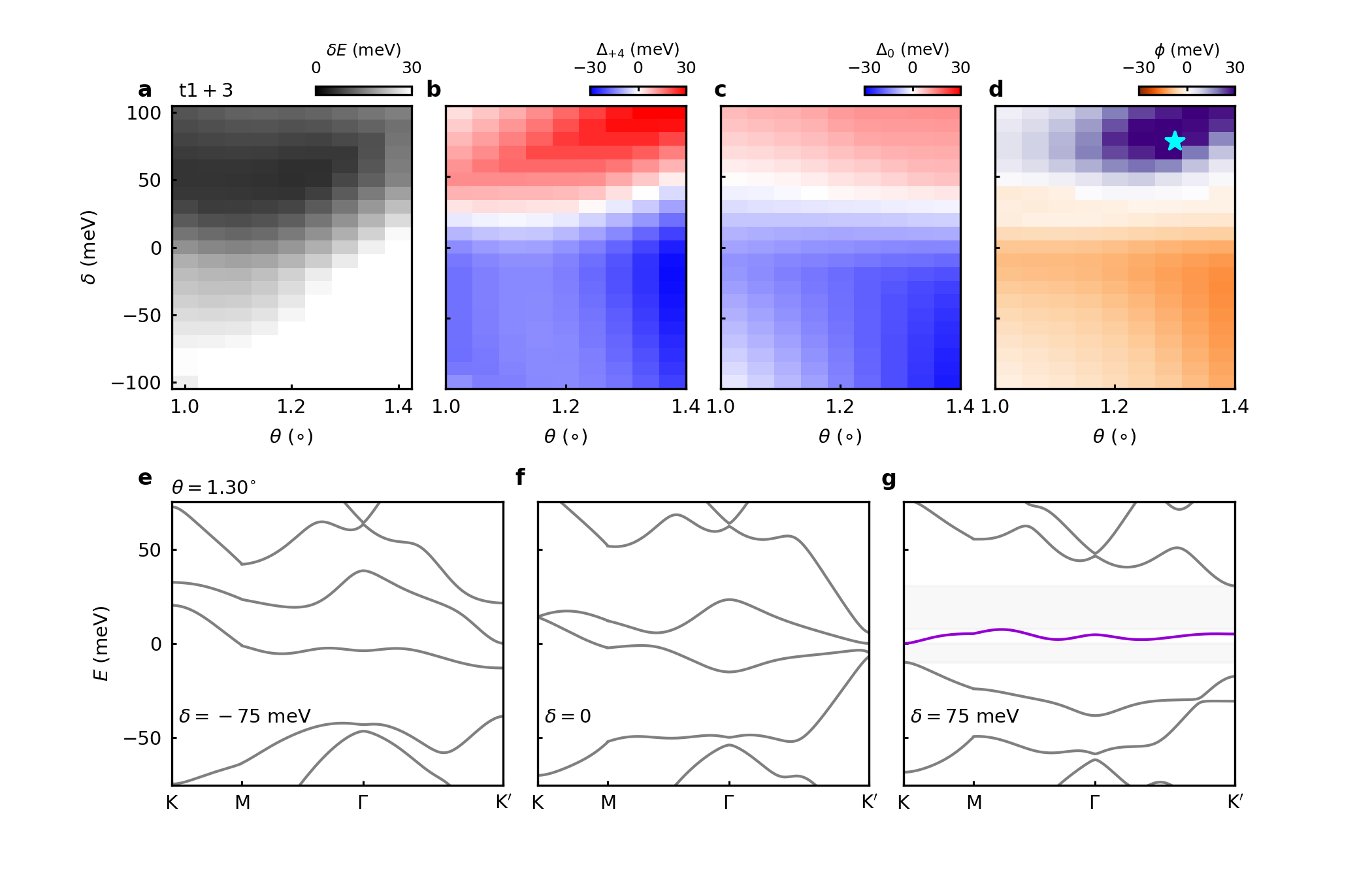} 
\caption{\textbf{Single particle calculations for t$\bm{1+3}$.} 
Similar to Supp. Fig. \ref{fig:phase_t12}. 
}
\label{fig:phase_t13}
\end{figure*}

\begin{figure*}[h]
\includegraphics[width=0.9\textwidth]{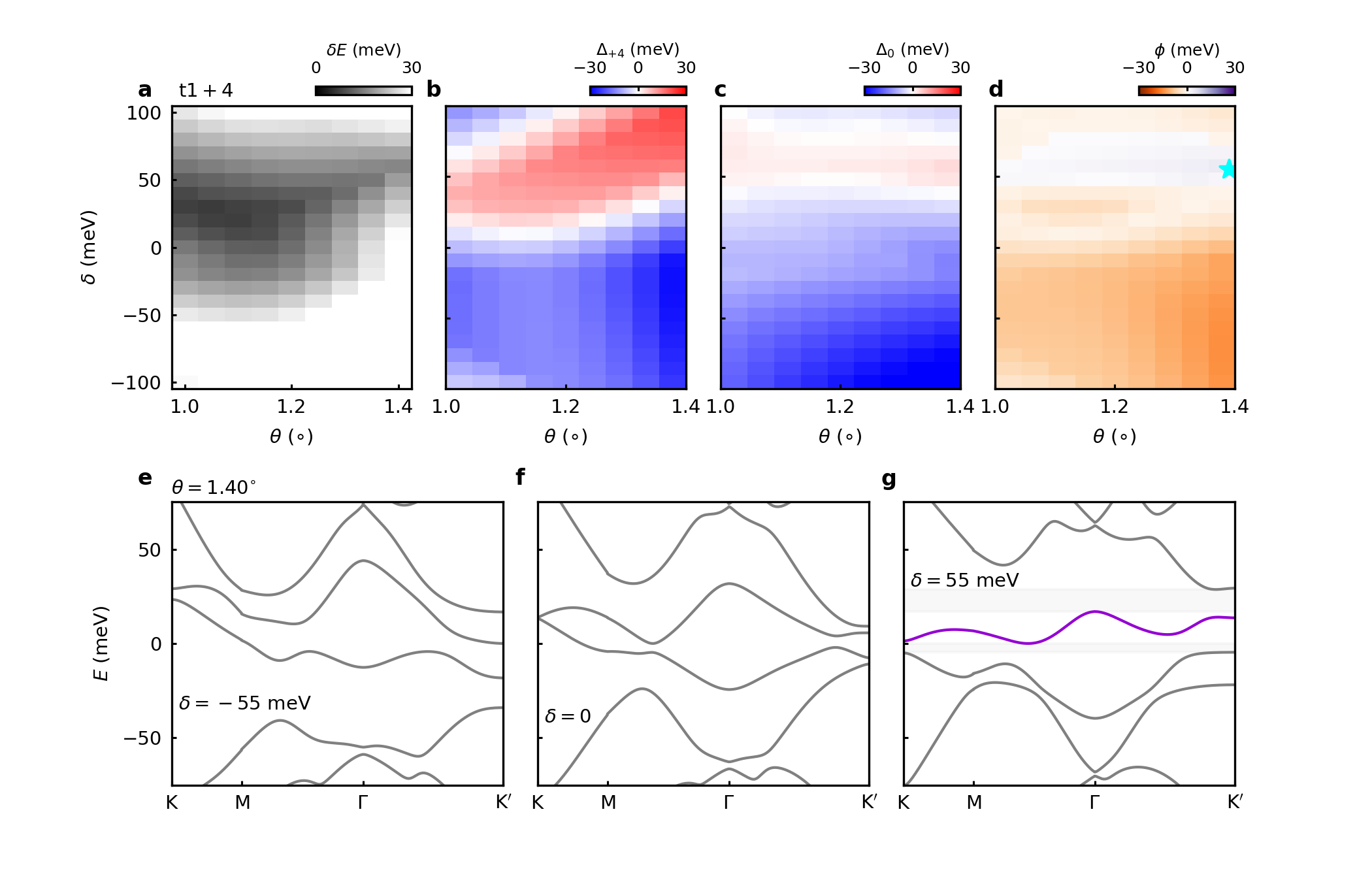} 
\caption{\textbf{Single particle calculations for t$\bm{1+4}$.} 
Similar to Supp. Fig. \ref{fig:phase_t12}. 
}
\label{fig:phase_t14}
\end{figure*}

\begin{figure*}[h]
\includegraphics[width=0.9\textwidth]{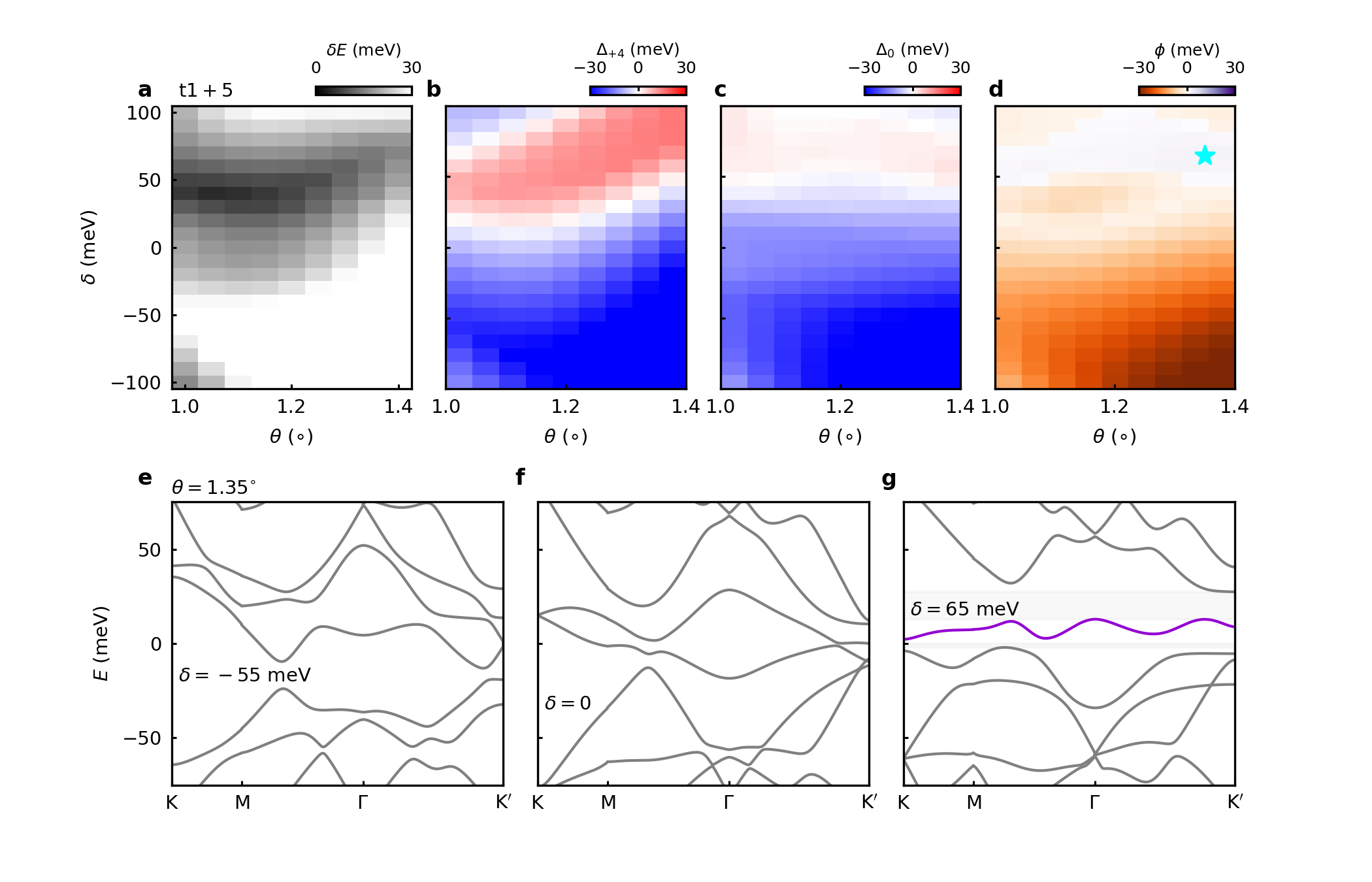} 
\caption{\textbf{Single particle calculations for t$\bm{1+5}$.} 
Similar to Supp. Fig. \ref{fig:phase_t12}. 
}
\label{fig:phase_t15}
\end{figure*}

\begin{figure*}[h]
\includegraphics[width=0.9\textwidth]{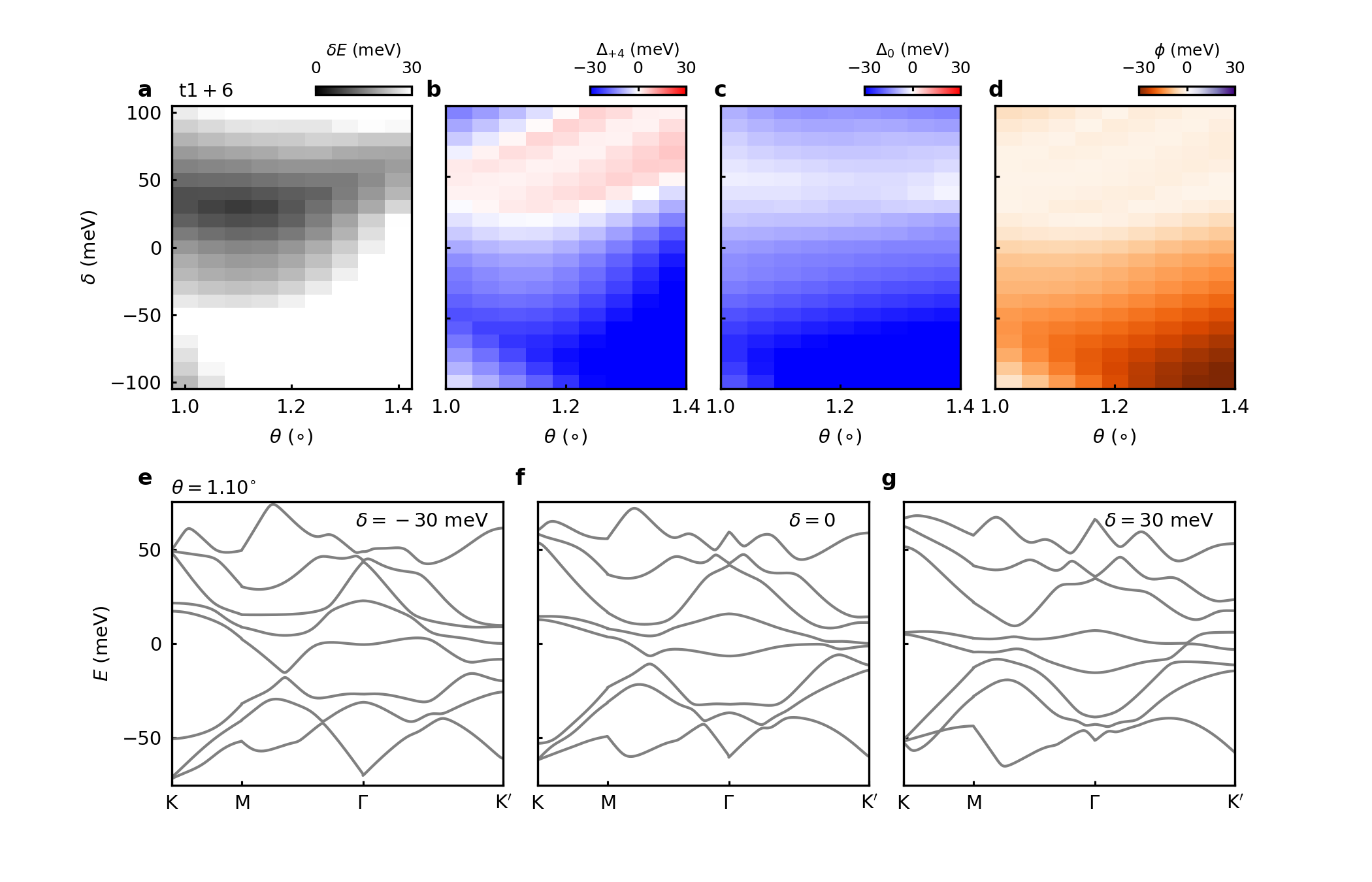} 
\caption{\textbf{Single particle calculations for t$\bm{1+6}$.} 
Similar to Supp. Fig. \ref{fig:phase_t12}. The \moire conduction band never becomes isolated for t$1+6$.
}
\label{fig:phase_t16}
\end{figure*}

\begin{figure*}[h]
\includegraphics[width=0.9\textwidth]{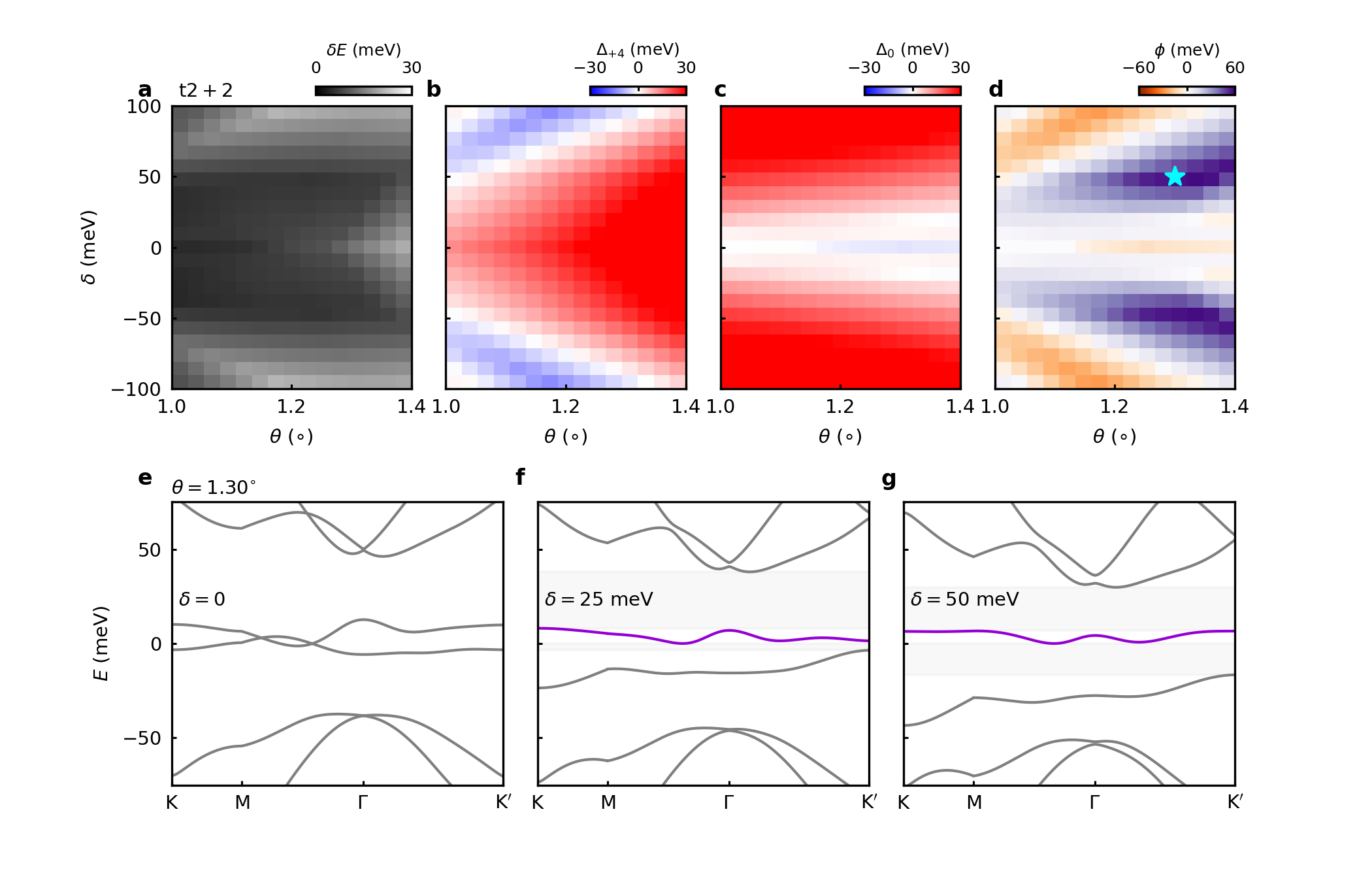} 
\caption{\textbf{Single particle calculations for t$\bm{2+2}$.} 
Similar to Supp. Fig. \ref{fig:phase_t12}. 
}
\label{fig:phase_t22}
\end{figure*}

\begin{figure*}[h]
\includegraphics[width=0.9\textwidth]{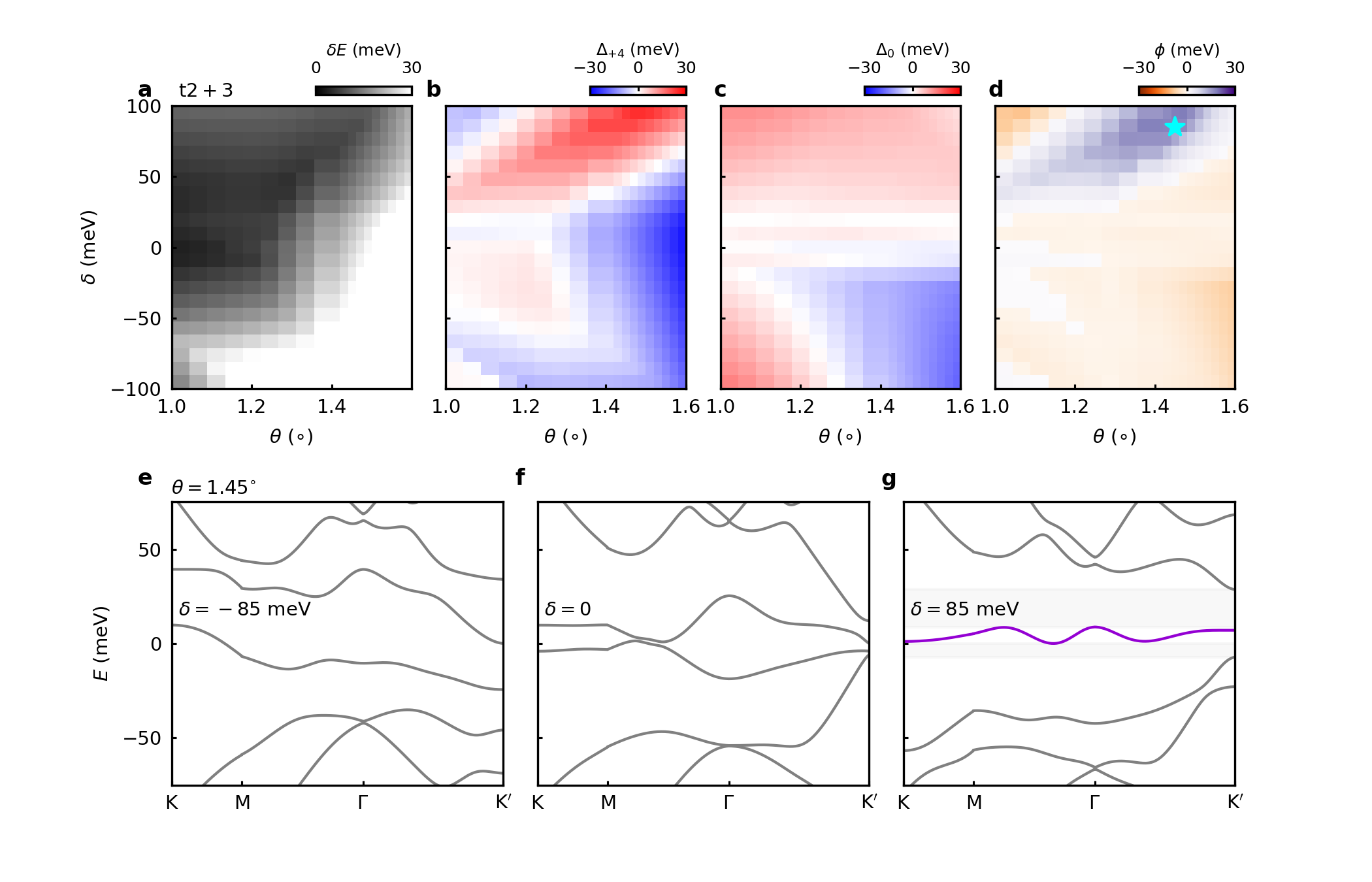} 
\caption{\textbf{Single particle calculations for t$\bm{2+3}$.} 
Similar to Supp. Fig. \ref{fig:phase_t12}. 
}
\label{fig:phase_t23}
\end{figure*}

\begin{figure*}[h]
\includegraphics[width=0.9\textwidth]{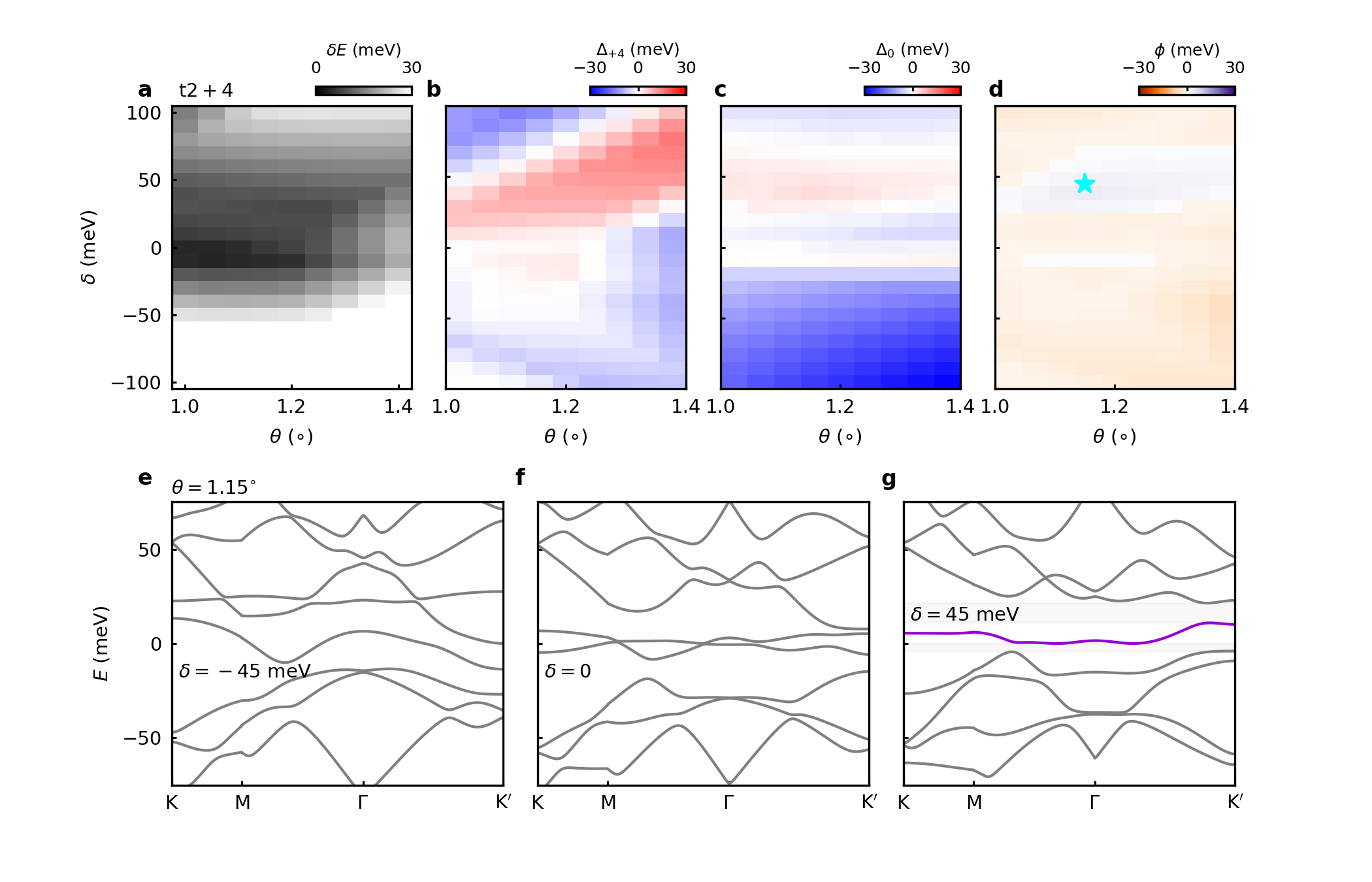} 
\caption{\textbf{Single particle calculations for t$\bm{2+4}$.} 
Similar to Supp. Fig. \ref{fig:phase_t12}. 
}
\label{fig:phase_t24}
\end{figure*}

\begin{figure*}[h]
\includegraphics[width=0.9\textwidth]{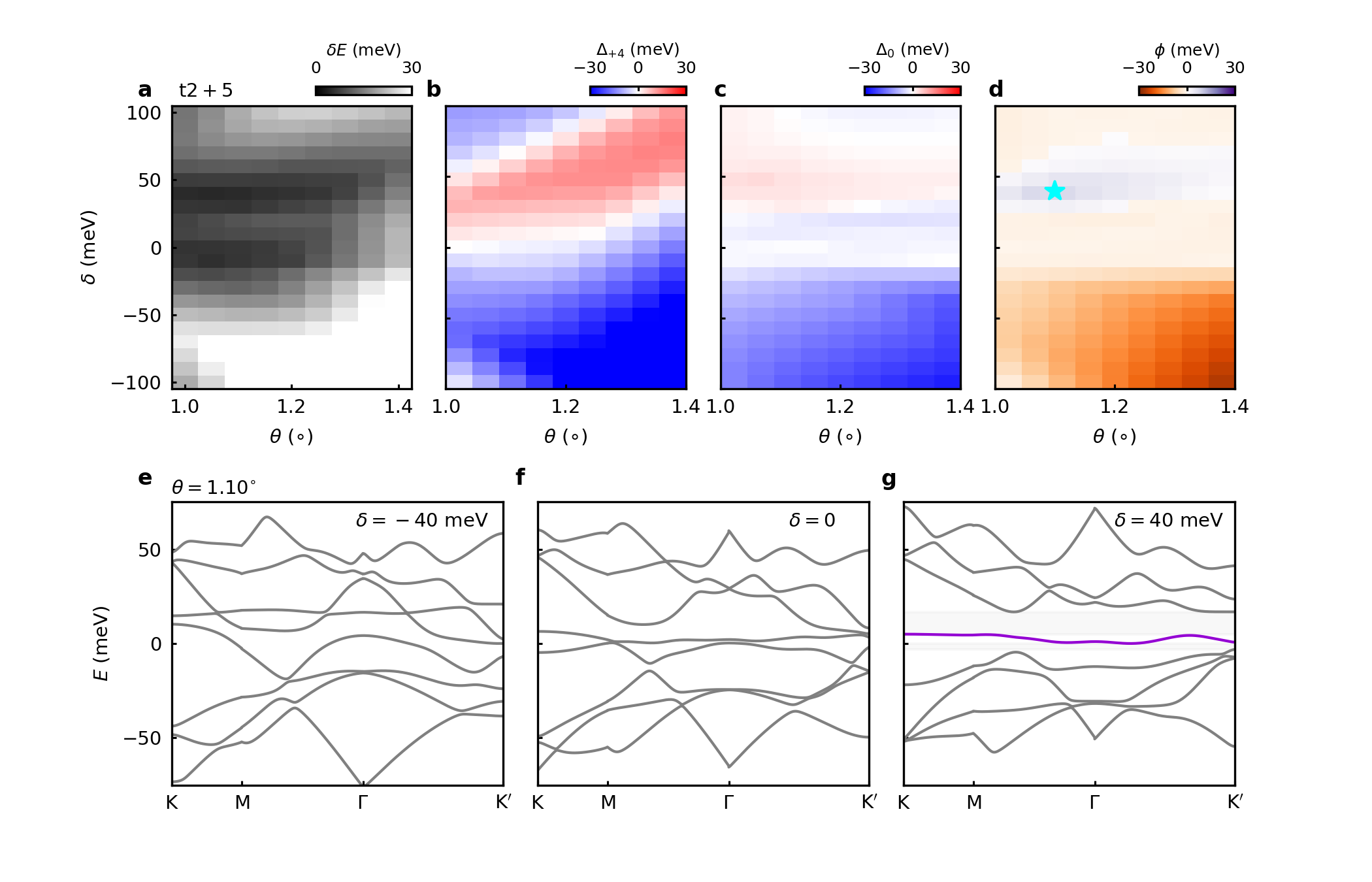} 
\caption{\textbf{Single particle calculations for t$\bm{2+5}$.} 
Similar to Supp. Fig. \ref{fig:phase_t12}. 
}
\label{fig:phase_t25}
\end{figure*}

\begin{figure*}[h]
\includegraphics[width=0.9\textwidth]{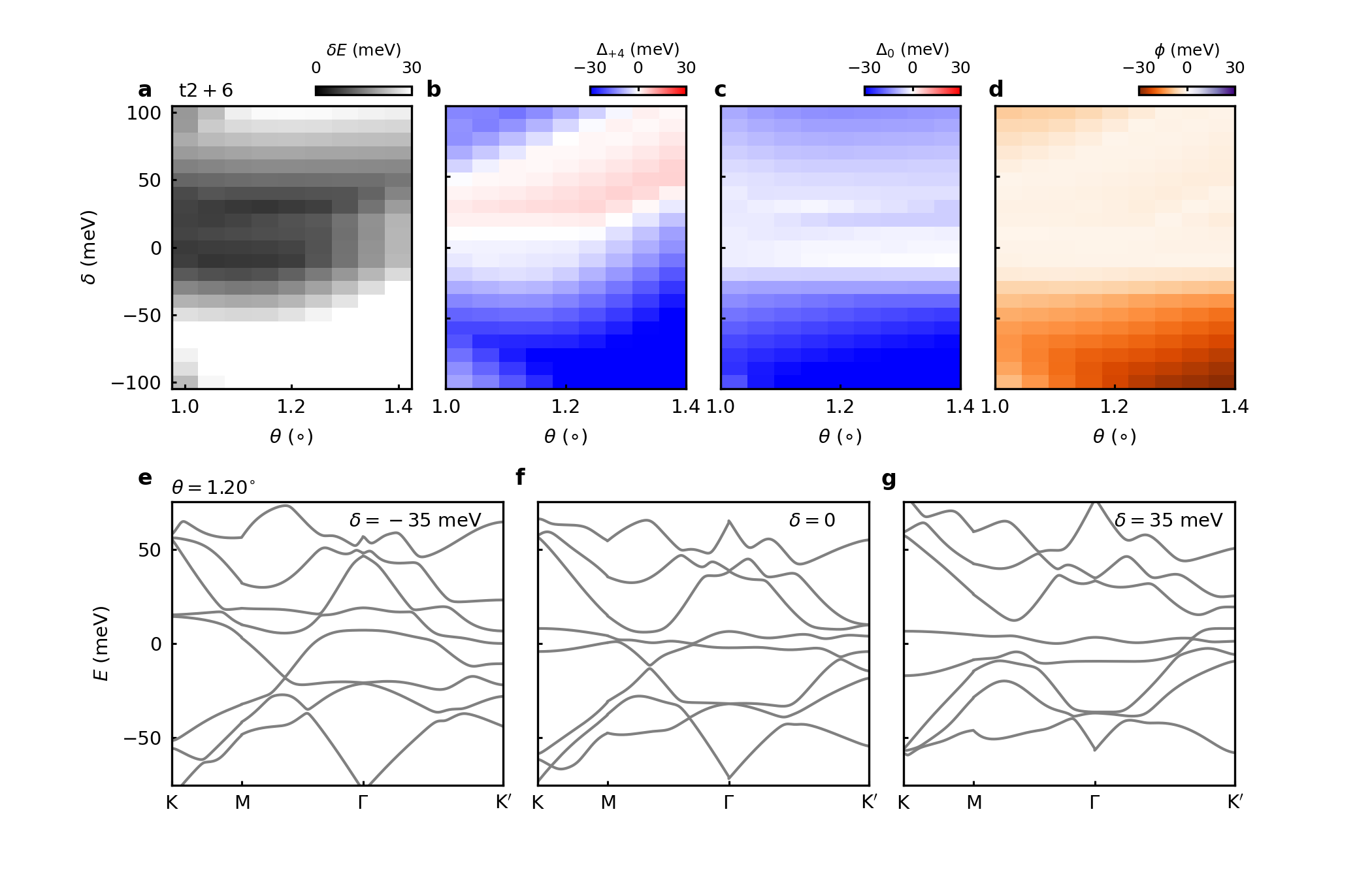} 
\caption{\textbf{Single particle calculations for t$\bm{2+6}$.} 
Similar to Supp. Fig. \ref{fig:phase_t12}. The \moire conduction band never becomes isolated for t$2+6$.
}
\label{fig:phase_t26}
\end{figure*}

\begin{figure*}[h]
\includegraphics[width=0.9\textwidth]{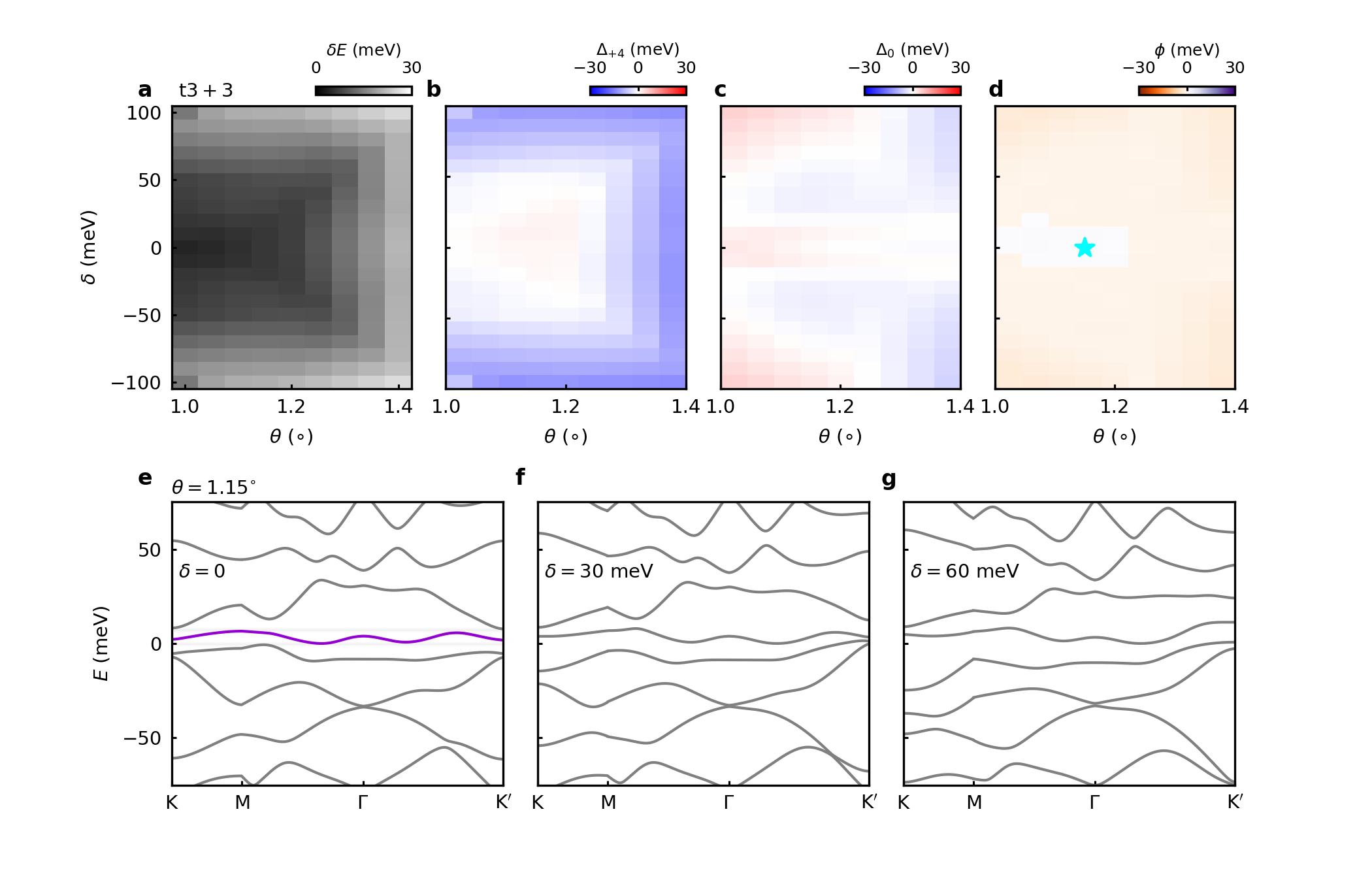} 
\caption{\textbf{Single particle calculations for t$\bm{3+3}$.} 
Similar to Supp. Fig. \ref{fig:phase_t12}. 
}
\label{fig:phase_t33}
\end{figure*}

\begin{figure*}[h]
\includegraphics[width=0.9\textwidth]{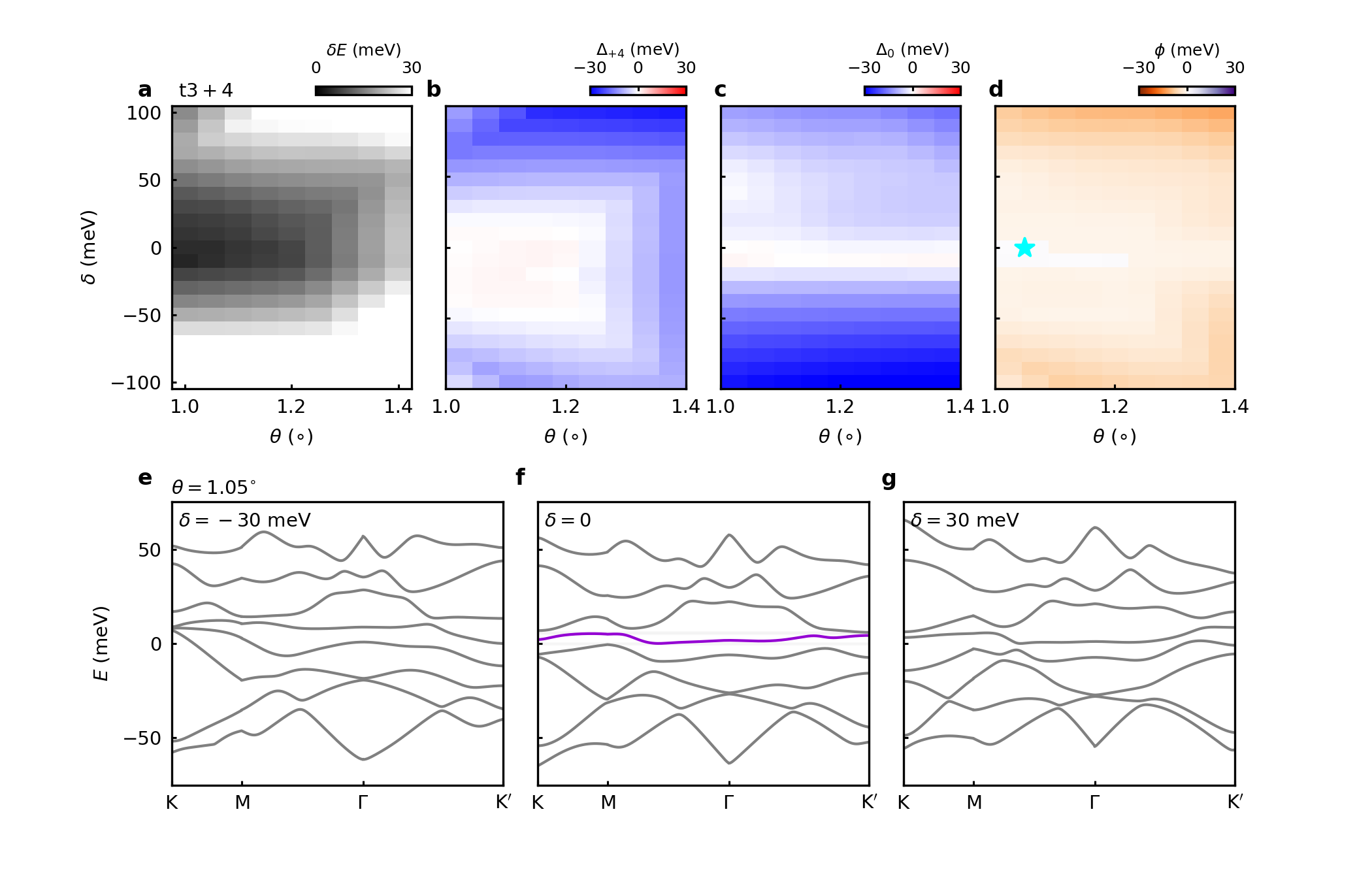} 
\caption{\textbf{Single particle calculations for t$\bm{3+4}$.} 
Similar to Supp. Fig. \ref{fig:phase_t12}. 
}
\label{fig:phase_t34}
\end{figure*}

\begin{figure*}[h]
\includegraphics[width=0.9\textwidth]{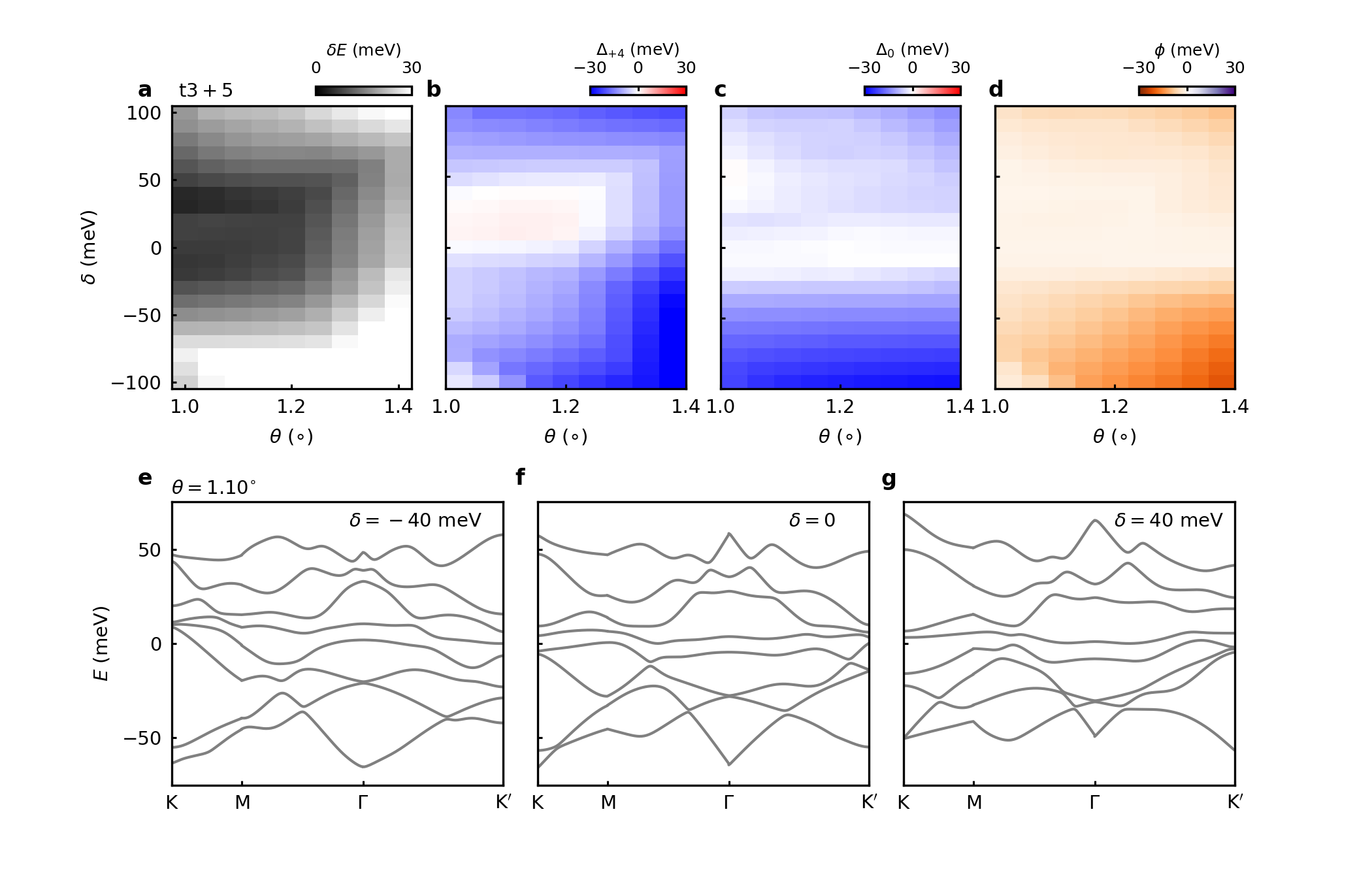} 
\caption{\textbf{Single particle calculations for t$\bm{3+5}$.} 
Similar to Supp. Fig. \ref{fig:phase_t12}. The \moire conduction band never becomes isolated for t$3+5$.
}
\label{fig:phase_t35}
\end{figure*}

\begin{figure*}[h]
\includegraphics[width=0.9\textwidth]{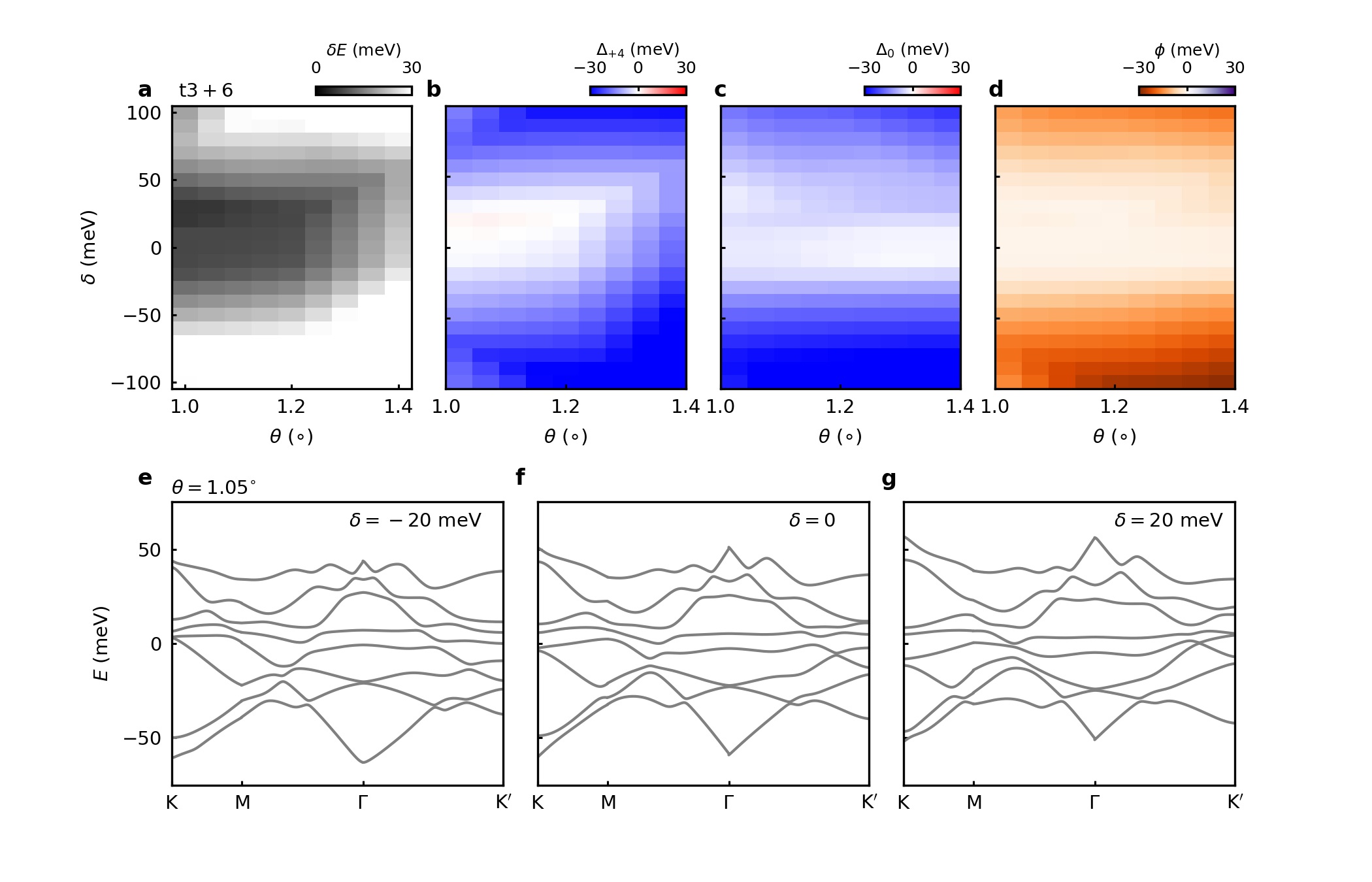} 
\caption{\textbf{Single particle calculations for t$\bm{3+6}$.} 
Similar to Supp. Fig. \ref{fig:phase_t12}. The \moire conduction band never becomes isolated for t$3+6$.
}
\label{fig:phase_t36}
\end{figure*}

\begin{figure*}[h]
\includegraphics[width=\textwidth]{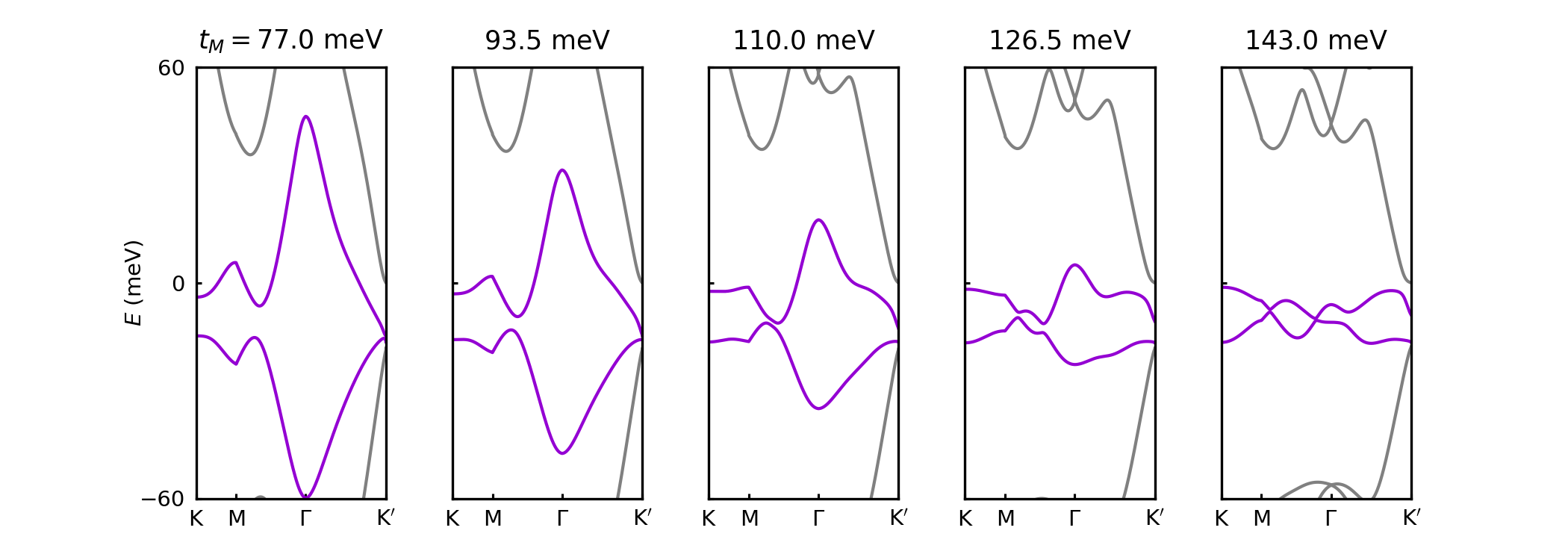} 
\caption{\textbf{Band structure calculations for the $\bm{\theta=1.50^{\circ}}$ t$\bm{2+3}$ at $\bm{\delta =0}$.}
Calculation results shown for varying values of the \moire coupling parameter $t_M$. The nominal value we used for all other calculations is $t_M=110\ \rm{meV}$ (center panel). Here both the \moire conduction and \moire valence band are shown in purple, to emphasize that the bands always overlap, inconsistent with our observation of an insulating state in Fig. 3d \& e of the main text. 
} 
\label{fig:bandstructures_D0}
\end{figure*}